\def\unredoffs{} \def\redoffs{\voffset=-.31truein\hoffset=-.59truein}
\def\speclscape{\special{ps: landscape}}
%
%%% DEC lw
%%%\def\unredoffs{} \def\redoffs{\voffset=-.40truein\hoffset=-.40truein}
%%%\def\speclscape{\special{landscape}}
%
%%% qms lasergrafix:
%\def\unredoffs{} \def\redoffs{\voffset=-.4truein\hoffset=.125truein}
%\def\speclscape{\special{qms: landscape}}
%
%%% saclay A4 paper:
%\def\unredoffs{\hoffset-.14truein\voffset-.2truein}
%\def\redoffs{\voffset=-.55truein\hoffset=-.1truein} \def\speclscape{}
%
%---------------------------------------------------------------------%
%
\newbox\leftpage \newdimen\fullhsize \newdimen\hstitle \newdimen\hsbody
\tolerance=1000\hfuzz=2pt
\catcode`@=11 % This allows us to modify PLAIN macros.
%
%	Big preprint format, 12pt type with normal size title at top of
% single column pages.
\def\PPTbig{%					% big preprint format
   \twelvepoint\unredoffs\baselineskip=16pt plus 2pt minus 1pt
   \hsbody=\hsize \hstitle=\hsize %take default values for unreduced format
}
%
%	Small preprint format, 10pt type in landscape double column
% format with large landscape cover page.
\def\PPTlittle{%
   \tenpoint\baselineskip=16pt plus 2pt minus 1pt \vsize=7truein
   \redoffs \hstitle=8truein\hsbody=4.75truein\fullhsize=10truein\hsize=\hsbody
   \output={\ifnum\pageno=0 %%% This is the HUTP version
     \shipout\vbox{\speclscape{\hsize\fullhsize\makeheadline}
     \hbox to \fullhsize{\hfill\pagebody\hfill}}\advancepageno
   \else
     \almostshipout{\leftline{\vbox{\pagebody\makefootline}}}\advancepageno
   \fi}
}
\def\almostshipout#1{\if L\l@r \count1=1 \message{[\the\count0.\the\count1]}
      \global\setbox\leftpage=#1 \global\let\l@r=R
 \else \count1=2
  \shipout\vbox{\speclscape{\hsize\fullhsize\makeheadline}
      \hbox to\fullhsize{\box\leftpage\hfil#1}}  \global\let\l@r=L\fi}
%---------------------------------------------------------------------
%
\newcount\yearltd\yearltd=\year\advance\yearltd by -1900

\def\Title#1#2{\nopagenumbers\abstractfont\hsize=\hstitle\rightline{#1}%
\vskip 1in\centerline{\titlefont #2}\abstractfont\vskip .5in\pageno=0}
%
% 	restores pagenumbers
%
%       use following instead of \Date on the preliminary draft,
%       puts date/time on each page in big mode, writes labels in margins

\def\draftmode{\message{ DRAFTMODE }\def\draftdate{{\rm preliminary draft:
\number\month/\number\day/\number\yearltd\ \ \hourmin}}%
\headline={\hfil\draftdate}\writelabels\baselineskip=20pt plus 2pt minus 2pt
 {\count255=\time\divide\count255 by 60 \xdef\hourmin{\number\count255}
  \multiply\count255 by-60\advance\count255 by\time
  \xdef\hourmin{\hourmin:\ifnum\count255<10 0\fi\the\count255}}}
%       use \nolabels to get rid of eqn, ref, and fig labels in draft mode
\def\nolabels{\def\wrlabeL##1{}\def\eqlabeL##1{}\def\reflabeL##1{}}
\def\writelabels{\def\wrlabeL##1{\leavevmode\vadjust{\rlap{\smash%
{\line{{\escapechar=` \hfill\rlap{\sevenrm\hskip.03in\string##1}}}}}}}%
\def\eqlabeL##1{{\escapechar-1\rlap{\sevenrm\hskip.05in\string##1}}}%
\def\reflabeL##1{\noexpand\llap{\noexpand\sevenrm\string\string\string##1}}}
\nolabels
%
% tagged sec numbers
\global\newcount\secno \global\secno=0
\global\newcount\meqno \global\meqno=1
\def\newsec#1{\global\advance\secno by1\message{(\the\secno. #1)}
%\ifx\answ\bigans \vfill\eject \else \bigbreak\bigskip \fi  %if desired
\global\subsecno=0\eqnres@t\noindent{\bf\the\secno. #1}
\writetoca{{\secsym} {#1}}\par\nobreak\medskip\nobreak}
\def\eqnres@t{\xdef\secsym{\the\secno.}\global\meqno=1\bigbreak\bigskip}
\def\sequentialequations{\def\eqnres@t{\bigbreak}}\xdef\secsym{}
\global\newcount\subsecno \global\subsecno=0
\def\subsec#1{\global\advance\subsecno by1\message{(\secsym\the\subsecno. #1)}
\ifnum\lastpenalty>9000\else\bigbreak\fi
\noindent{\it\secsym\the\subsecno. #1}\writetoca{\string\quad
{\secsym\the\subsecno.} {#1}}\par\nobreak\medskip\nobreak}
\def\appendix#1#2{\global\meqno=1\global\subsecno=0\xdef\secsym{\hbox{#1.}}
\bigbreak\bigskip\noindent{\bf Appendix #1. #2}\message{(#1. #2)}
\writetoca{Appendix {#1.} {#2}}\par\nobreak\medskip\nobreak}
%
%       \eqn\label{a+b=c}	gives displayed equation, numbered
%				consecutively within sections.
%     \eqnn and \eqna define labels in advance (of eqalign?)
%
\def\eqnn#1{\xdef #1{(\secsym\the\meqno)}\writedef{#1\leftbracket#1}%
\global\advance\meqno by1\wrlabeL#1}
\def\eqna#1{\xdef #1##1{\hbox{$(\secsym\the\meqno##1)$}}
\writedef{#1\numbersign1\leftbracket#1{\numbersign1}}%
\global\advance\meqno by1\wrlabeL{#1$\{\}$}}
\def\eqn#1#2{\xdef #1{(\secsym\the\meqno)}\writedef{#1\leftbracket#1}%
\global\advance\meqno by1$$#2\eqno#1\eqlabeL#1$$}
%
%			 footnotes
\newskip\footskip\footskip14pt plus 1pt minus 1pt %sets footnote baselineskip
\def\footnotefont{\ninepoint}\def\f@t#1{\footnotefont #1\@foot}
\def\f@@t{\baselineskip\footskip\bgroup\footnotefont\aftergroup\@foot\let\next}
\setbox\strutbox=\hbox{\vrule height9.5pt depth4.5pt width0pt}
\global\newcount\ftno \global\ftno=0
\def\foot{\global\advance\ftno by1\footnote{$^{\the\ftno}$}}
%
%say \footend to put footnotes at end
%will cause problems if \ref used inside \foot, instead use \nref before
\newwrite\ftfile
\def\footend{\def\foot{\global\advance\ftno by1\chardef\wfile=\ftfile
$^{\the\ftno}$\ifnum\ftno=1\immediate\openout\ftfile=foots.tmp\fi%
\immediate\write\ftfile{\noexpand\smallskip%
\noexpand\item{f\the\ftno:\ }\pctsign}\findarg}%
\def\footatend{\vfill\eject\immediate\closeout\ftfile{\parindent=20pt
\centerline{\bf Footnotes}\nobreak\bigskip\input foots.tmp }}}
\def\footatend{}
%
%     \ref\label{text}
% generates a number, assigns it to \label, generates an entry.
% To list the refs on a separate page,  \listrefs
%
\global\newcount\refno \global\refno=1
\newwrite\rfile
\def\ref{[\the\refno]\nref}
\def\nref#1{\xdef#1{[\the\refno]}\writedef{#1\leftbracket#1}%
\ifnum\refno=1\immediate\openout\rfile=refs.tmp\fi
\global\advance\refno by1\chardef\wfile=\rfile\immediate
\write\rfile{\noexpand\item{#1\ }\reflabeL{#1\hskip.31in}\pctsign}\findarg}
%	horrible hack to sidestep tex \write limitation
\def\findarg#1#{\begingroup\obeylines\newlinechar=`\^^M\pass@rg}
{\obeylines\gdef\pass@rg#1{\writ@line\relax #1^^M\hbox{}^^M}%
\gdef\writ@line#1^^M{\expandafter\toks0\expandafter{\striprel@x #1}%
\edef\next{\the\toks0}\ifx\next\em@rk\let\next=\endgroup\else\ifx\next\empty%
\else\immediate\write\wfile{\the\toks0}\fi\let\next=\writ@line\fi\next\relax}}
\def\striprel@x#1{} \def\em@rk{\hbox{}}
\def\lref{\begingroup\obeylines\lr@f}
\def\lr@f#1#2{\gdef#1{\ref#1{#2}}\endgroup\unskip}
\def\semi{;\hfil\break}
\def\addref#1{\immediate\write\rfile{\noexpand\item{}#1}} %now unnecessary
\def\listrefs{\footatend\vfill\supereject\immediate\closeout\rfile\writestoppt
\baselineskip=14pt\centerline{{\bf References}}\bigskip{\frenchspacing%
\parindent=20pt\escapechar=` \input refs.tmp\vfill\eject}\nonfrenchspacing}
\def\startrefs#1{\immediate\openout\rfile=refs.tmp\refno=#1}
\def\xref{\expandafter\xr@f}\def\xr@f[#1]{#1}
\def\refs#1{\count255=1[\r@fs #1{\hbox{}}]}
\def\r@fs#1{\ifx\und@fined#1\message{reflabel \string#1 is undefined.}%
\nref#1{need to supply reference \string#1.}\fi%
\vphantom{\hphantom{#1}}\edef\next{#1}\ifx\next\em@rk\def\next{}%
\else\ifx\next#1\ifodd\count255\relax\xref#1\count255=0\fi%
\else#1\count255=1\fi\let\next=\r@fs\fi\next}
%

%
% this is ugly, but moore insists
\newwrite\ffile\global\newcount\figno \global\figno=1
\def\fig{fig.~\the\figno\nfig}
\def\nfig#1{\xdef#1{fig.~\the\figno}%
\writedef{#1\leftbracket fig.\noexpand~\the\figno}%
\ifnum\figno=1\immediate\openout\ffile=figs.tmp\fi\chardef\wfile=\ffile%
\immediate\write\ffile{\noexpand\medskip\noexpand\item{Fig.\ \the\figno. }
\reflabeL{#1\hskip.55in}\pctsign}\global\advance\figno by1\findarg}
\def\xfig{\expandafter\xf@g}\def\xf@g fig.\penalty\@M\ {}
\def\figs#1{figs.~\f@gs #1{\hbox{}}}
\def\f@gs#1{\edef\next{#1}\ifx\next\em@rk\def\next{}\else
\ifx\next#1\xfig #1\else#1\fi\let\next=\f@gs\fi\next}
\newwrite\lfile
{\escapechar-1\xdef\pctsign{\string\%}\xdef\leftbracket{\string\{}
\xdef\rightbracket{\string\}}\xdef\numbersign{\string\#}}
\def\writedefs{\immediate\openout\lfile=labeldefs.tmp \def\writedef##1{%
\immediate\write\lfile{\string\def\string##1\rightbracket}}}
%	\readdefs read the file written by \writedefs
\newread\lfilein
		% and input it
%
\def\writestop{\def\writestoppt{\immediate\write\lfile{\string\pageno%
\the\pageno\string\startrefs\leftbracket\the\refno\rightbracket%
\string\def\string\secsym\leftbracket\secsym\rightbracket%
\string\secno\the\secno\string\meqno\the\meqno}\immediate\closeout\lfile}}
\def\writestoppt{}\def\writedef#1{}
\def\seclab#1{\xdef #1{\the\secno}\writedef{#1\leftbracket#1}\wrlabeL{#1=#1}}
\def\subseclab#1{\xdef #1{\secsym\the\subsecno}%
\writedef{#1\leftbracket#1}\wrlabeL{#1=#1}}
\newwrite\tfile \def\writetoca#1{}
\def\leaderfill{\leaders\hbox to 1em{\hss.\hss}\hfill}
%	use this to write file with table of contents
\def\writetoc{\immediate\openout\tfile=toc.tmp
   \def\writetoca##1{{\edef\next{\write\tfile{\noindent ##1
   \string\leaderfill {\noexpand\number\pageno} \par}}\next}}}
%       and this lists table of contents on second pass

%
%---------------------------------------------------------------------
%	Unpleasantness in calling in abstract and title fonts
%	This is now less unpleasant -- the unpleasantness has been
% generalized below.

\def\titlefont{%				% title font
   \ifx\answ\bigans \sixteenpoint%		% big is 16pt
      \else\twentypoint\fi}%			% small is 20pt cover

\global\font\tencsc=cmcsc10
\global\font\twelvecsc=cmcsc10 scaled \magstep1

\def\authorfont{%				% author font
   \ifx\answ\bigans\tenpoint\tencsc%		% big is 10pt
   \else\twelvepoint\twelvecsc\fi}		% small is 12pt cover

\def\abstractfont{%				% abstract font
   \ifx\answ\bigans\tenpoint%			% big is 10pt
   \else\twelvepoint\fi}			% small is 12pt cover

%---------------------------------------------------------------------
%

\hyphenation{anom-aly anom-alies coun-ter-term coun-ter-terms}
\def\inv{^{\raise.15ex\hbox{${\scriptscriptstyle -}$}\kern-.05em 1}}

\def\Dsl{\,\raise.15ex\hbox{/}\mkern-13.5mu D} %this one can be subscripted
\def\dsl{\raise.15ex\hbox{/}\kern-.57em\partial}

 %pound sterling
\def\lspace{\ifx\answ\bigans{}\else\qquad\fi}
\def\lbspace{\ifx\answ\bigans{}\else\hskip-.2in\fi} % $$\lbspace...$$
\def\boxeqn#1{\vcenter{\vbox{\hrule\hbox{\vrule\kern3pt\vbox{\kern3pt
	\hbox{${\displaystyle #1}$}\kern3pt}\kern3pt\vrule}\hrule}}}
\def\mbox#1#2{\vcenter{\hrule \hbox{\vrule height#2in
		\kern#1in \vrule} \hrule}}  %e.g. \mbox{.1}{.1}
%	matters of taste
%\def\tilde{\widetilde} \def\bar{\overline} \def\hat{\widehat}
%
% some sample definitions
  %     curly letters
   \def\CG{{\cal G}}
\def\CL{{\cal L}}   \def\CU{{\cal U}}
  \def\CD{{\cal D}}

\def\darr#1{\raise1.5ex\hbox{$\leftrightarrow$}\mkern-16.5mu #1}
 %pound sterling

\def\half{{\textstyle{1\over2}}} %puts a small half in a displayed eqn
\def\roughly#1{\raise.3ex\hbox{$#1$\kern-.75em\lower1ex\hbox{$\sim$}}}

%======================================================================*
%% file: TXSfonts.tex                              TeXsis version 2.15
%  $Revision: 16.0 $  :  $Date: 1992/08/08 22:20:36 $  :  $Author: myers $
%======================================================================*
\catcode`@=11                           % @ is a letter here
\newskip\ttglue

%========================================================================
% FONT DEFINITIONS
%       Macros to define fonts as they are used. Each redefines itself to be
% \relax after it is first called so that it is disabled.

%%++ 9 pt fonts:
\def\ninefonts{%
   \global\font\ninerm=cmr9
   \global\font\ninei=cmmi9
   \global\font\ninesy=cmsy9
   \global\font\nineex=cmex10
   \global\font\ninebf=cmbx9
   \global\font\ninesl=cmsl9
   \global\font\ninett=cmtt9
   \global\font\nineit=cmti9
   \skewchar\ninei='177
   \skewchar\ninesy='60
   \hyphenchar\ninett=-1
   \moreninefonts                               % any custom fonts
   \gdef\ninefonts{\relax}}

\def\moreninefonts{\relax}                      % User customization hook

%%++ 10 pt fonts:                               % most alread loaded by Plain
                              % 10 pt sans serif

                       % customization hook

%%++ 11 pt fonts:
\def\elevenfonts{%
   \global\font\elevenrm=cmr10 scaled \magstephalf
   \global\font\eleveni=cmmi10 scaled \magstephalf
   \global\font\elevensy=cmsy10 scaled \magstephalf
   \global\font\elevenex=cmex10
   \global\font\elevenbf=cmbx10 scaled \magstephalf
   \global\font\elevensl=cmsl10 scaled \magstephalf
   \global\font\eleventt=cmtt10 scaled \magstephalf
   \global\font\elevenit=cmti10 scaled \magstephalf
   \global\font\elevenss=cmss10 scaled \magstephalf
   \skewchar\eleveni='177%%
   \skewchar\elevensy='60%%
   \hyphenchar\eleventt=-1%%
   \moreelevenfonts                            % any other custom fonts
   \gdef\elevenfonts{\relax}}%

\def\moreelevenfonts{\relax}                    % User customization hook

%%++ 12 pt fonts:
\def\twelvefonts{%  initialize 12pt fonts
   \global\font\twelverm=cmr10 scaled \magstep1%%
   \global\font\twelvei=cmmi10 scaled \magstep1%%
   \global\font\twelvesy=cmsy10 scaled \magstep1%%
   \global\font\twelveex=cmex10 scaled \magstep1%%
   \global\font\twelvebf=cmbx10 scaled \magstep1%%
   \global\font\twelvesl=cmsl10 scaled \magstep1%%
   \global\font\twelvett=cmtt10 scaled \magstep1%%
   \global\font\twelveit=cmti10 scaled \magstep1%%
   \global\font\twelvess=cmss10 scaled \magstep1%%
   \skewchar\twelvei='177%%
   \skewchar\twelvesy='60%%
   \hyphenchar\twelvett=-1%%
   \moretwelvefonts                             % any other custom fonts
   \gdef\twelvefonts{\relax}}

\def\moretwelvefonts{\relax}                    % User customization hook

%%++ 14 pt fonts:
\def\fourteenfonts{%
   \global\font\fourteenrm=cmr10 scaled \magstep2%%
   \global\font\fourteeni=cmmi10 scaled \magstep2%%
   \global\font\fourteensy=cmsy10 scaled \magstep2%%
   \global\font\fourteenex=cmex10 scaled \magstep2%%
   \global\font\fourteenbf=cmbx10 scaled \magstep2%%
   \global\font\fourteensl=cmsl10 scaled \magstep2%%
   \global\font\fourteenit=cmti10 scaled \magstep2%%
   \global\font\fourteenss=cmss10 scaled \magstep2%%
   \skewchar\fourteeni='177%%
   \skewchar\fourteensy='60%%
   \morefourteenfonts                           % any custom fonts
   \gdef\fourteenfonts{\relax}}

\def\morefourteenfonts{\relax}                  % customization hook

%%++ 16 pt fonts:
\def\sixteenfonts{%
   \global\font\sixteenrm=cmr10 scaled \magstep3%%
   \global\font\sixteeni=cmmi10 scaled \magstep3%%
   \global\font\sixteensy=cmsy10 scaled \magstep3%%
   \global\font\sixteenex=cmex10 scaled \magstep3%%
   \global\font\sixteenbf=cmbx10 scaled \magstep3%%
   \global\font\sixteensl=cmsl10 scaled \magstep3%%
   \global\font\sixteenit=cmti10 scaled \magstep3%%
   \skewchar\sixteeni='177%%
   \skewchar\sixteensy='60%%
   \moresixteenfonts                            % any custom fonts
   \gdef\sixteenfonts{\relax}}

\def\moresixteenfonts{\relax}                   % User customization hook

%%++ 20 pt fonts:
\def\twentyfonts{%
   \global\font\twentyrm=cmr10 scaled \magstep4%%
   \global\font\twentyi=cmmi10 scaled \magstep4%%
   \global\font\twentysy=cmsy10 scaled \magstep4%%
   \global\font\twentyex=cmex10 scaled \magstep4%%
   \global\font\twentybf=cmbx10 scaled \magstep4%%
   \global\font\twentysl=cmsl10 scaled \magstep4%%
   \global\font\twentyit=cmti10 scaled \magstep4%%
   \skewchar\twentyi='177%%
   \skewchar\twentysy='60%%
   \moretwentyfonts                             % any custom fonts
   \gdef\twentyfonts{\relax}}

\def\moretwentyfonts{\relax}                    % User customization hook

%%++ 24 pt fonts:
\def\twentyfourfonts{%
   \global\font\twentyfourrm=cmr10 scaled \magstep5%%
   \global\font\twentyfouri=cmmi10 scaled \magstep5%%
   \global\font\twentyfoursy=cmsy10 scaled \magstep5%%
   \global\font\twentyfourex=cmex10 scaled \magstep5%%
   \global\font\twentyfourbf=cmbx10 scaled \magstep5%%
   \global\font\twentyfoursl=cmsl10 scaled \magstep5%%
   \global\font\twentyfourit=cmti10 scaled \magstep5%%
   \skewchar\twentyfouri='177%%
   \skewchar\twentyfoursy='60%%
   \moretwentyfourfonts                         % any custom fonts
   \gdef\twentyfourfonts{\relax}}

\def\moretwentyfourfonts{\relax}                % User customization hook

%========================================================================
% MATH ITALIC BOLD.
%     This makes boldfaced Greek characters for titles. Available only
% with cm fonts, generally only in 10pt and larger. Hence we use standard
% math fonts for the smaller sizes.

%     Macros to load math italic bold fonts:

\def\tenmibfonts{%                      % ten pt math italic bold
   \global\font\tenmib=cmmib10
   \global\font\tenbsy=cmbsy10
   \skewchar\tenmib='177%%
   \skewchar\tenbsy='60%%
   \gdef\tenmibfonts{\relax}}

\def\elevenmibfonts{%                   % eleven pt math italic bold
   \global\font\elevenmib=cmmib10 scaled \magstephalf
   \global\font\elevenbsy=cmbsy10 scaled \magstephalf
   \skewchar\elevenmib='177%%
   \skewchar\elevenbsy='60%%
   \gdef\elevenmibfonts{\relax}}

\def\twelvemibfonts{%                   % twelve pt math italic bold
   \global\font\twelvemib=cmmib10 scaled \magstep1%%
   \global\font\twelvebsy=cmbsy10 scaled \magstep1%%
   \skewchar\twelvemib='177%%
   \skewchar\twelvebsy='60%%
   \gdef\twelvemibfonts{\relax}}

\def\fourteenmibfonts{%                 % fourteen pt math italic bold
   \global\font\fourteenmib=cmmib10 scaled \magstep2%%
   \global\font\fourteenbsy=cmbsy10 scaled \magstep2%%
   \skewchar\fourteenmib='177%%
   \skewchar\fourteenbsy='60%%
   \gdef\fourteenmibfonts{\relax}}

\def\sixteenmibfonts{%                  % sixteen pt math italic bold
   \global\font\sixteenmib=cmmib10 scaled \magstep3%%
   \global\font\sixteenbsy=cmbsy10 scaled \magstep3%%
   \skewchar\sixteenmib='177%%
   \skewchar\sixteenbsy='60%%
   \gdef\sixteenmibfonts{\relax}}

\def\twentymibfonts{%                   % twenty pt math italic bold
   \global\font\twentymib=cmmib10 scaled \magstep4%%
   \global\font\twentybsy=cmbsy10 scaled \magstep4%%
   \skewchar\twentymib='177%%
   \skewchar\twentybsy='60%%
   \gdef\twentymibfonts{\relax}}

\def\twentyfourmibfonts{%               % twentyfour pt math italic bold
   \global\font\twentyfourmib=cmmib10 scaled \magstep5%%
   \global\font\twentyfourbsy=cmbsy10 scaled \magstep5%%
   \skewchar\twentyfourmib='177%%
   \skewchar\twentyfourbsy='60%%
   \gdef\twentyfourmibfonts{\relax}}

%     \mib changes to math italic bold by setting \textfont1 and \textfont2
% to cmmib10 and cmbsy10 respectively. This gets redefined by all the size
% changing macros.

\def\mib{%    set math italic bold for use in $...$
   \tenmibfonts
   \textfont0=\tenbf\scriptfont0=\sevenbf
   \scriptscriptfont0=\fivebf
   \textfont1=\tenmib\scriptfont1=\seveni
   \scriptscriptfont1=\fivei
   \textfont2=\tenbsy\scriptfont2=\sevensy
   \scriptscriptfont2=\fivesy}

%========================================================================
% Dummy SCR fonts
\def\scrfonts{\relax}

%========================================================================
% FONT SIZES
%       Macros to change font sizes. Each of these load fonts the first
% time it is called.

%       Switch to 9 point type
\def\ninepoint{\ninefonts               % load  9pt fonts if needed
   \def\rm{\fam0\ninerm}%
   \textfont0=\ninerm\scriptfont0=\sevenrm\scriptscriptfont0=\fiverm
   \textfont1=\ninei\scriptfont1=\seveni\scriptscriptfont1=\fivei
   \textfont2=\ninesy\scriptfont2=\sevensy\scriptscriptfont2=\fivesy
   \textfont3=\nineex\scriptfont3=\nineex\scriptscriptfont3=\nineex
   \textfont\itfam=\nineit\def\it{\fam\itfam\nineit}%
   \textfont\slfam=\ninesl\def\sl{\fam\slfam\ninesl}%
   \textfont\ttfam=\ninett\def\tt{\fam\ttfam\ninett}%
   \textfont\bffam=\ninebf
   \scriptfont\bffam=\sevenbf
   \scriptscriptfont\bffam=\fivebf\def\bf{\fam\bffam\ninebf}%
   \def\mib{\relax}%
   \def\scr{\relax}%
   \tt\ttglue=.5emplus.25emminus.15em
   \normalbaselineskip=11pt
   \setbox\strutbox=\hbox{\vrule height 8pt depth 3pt width 0pt}%
   \normalbaselines\rm\singlespaced}%

%       Switch to 10 point type
\def\tenpoint{%                         % 10pt already loaded by default
   \def\rm{\fam0\tenrm}%
   \textfont0=\tenrm\scriptfont0=\sevenrm\scriptscriptfont0=\fiverm
   \textfont1=\teni\scriptfont1=\seveni\scriptscriptfont1=\fivei
   \textfont2=\tensy\scriptfont2=\sevensy\scriptscriptfont2=\fivesy
   \textfont3=\tenex\scriptfont3=\tenex\scriptscriptfont3=\tenex
   \textfont\itfam=\tenit\def\it{\fam\itfam\tenit}%
   \textfont\slfam=\tensl\def\sl{\fam\slfam\tensl}%
   \textfont\ttfam=\tentt\def\tt{\fam\ttfam\tentt}%
   \textfont\bffam=\tenbf
   \scriptfont\bffam=\sevenbf
   \scriptscriptfont\bffam=\fivebf\def\bf{\fam\bffam\tenbf}%
   \def\mib{%
      \tenmibfonts
      \textfont0=\tenbf\scriptfont0=\sevenbf
      \scriptscriptfont0=\fivebf
      \textfont1=\tenmib\scriptfont1=\seveni
      \scriptscriptfont1=\fivei
      \textfont2=\tenbsy\scriptfont2=\sevensy
      \scriptscriptfont2=\fivesy}%
   \def\scr{\scrfonts
      \global\textfont\scrfam=\tenscr\fam\scrfam\tenscr}%
   \tt\ttglue=.5emplus.25emminus.15em
   \normalbaselineskip=12pt
   \setbox\strutbox=\hbox{\vrule height 8.5pt depth 3.5pt width 0pt}%
   \normalbaselines\rm\singlespaced}%

%       Switch to 11 point type
\def\elevenpoint{\elevenfonts           % load 11pt fonts if needed
   \def\rm{\fam0\elevenrm}%
   \textfont0=\elevenrm\scriptfont0=\sevenrm\scriptscriptfont0=\fiverm
   \textfont1=\eleveni\scriptfont1=\seveni\scriptscriptfont1=\fivei
   \textfont2=\elevensy\scriptfont2=\sevensy\scriptscriptfont2=\fivesy
   \textfont3=\elevenex\scriptfont3=\elevenex\scriptscriptfont3=\elevenex
   \textfont\itfam=\elevenit\def\it{\fam\itfam\elevenit}%
   \textfont\slfam=\elevensl\def\sl{\fam\slfam\elevensl}%
   \textfont\ttfam=\eleventt\def\tt{\fam\ttfam\eleventt}%
   \textfont\bffam=\elevenbf
   \scriptfont\bffam=\sevenbf
   \scriptscriptfont\bffam=\fivebf\def\bf{\fam\bffam\elevenbf}%
   \def\mib{%
      \elevenmibfonts
      \textfont0=\elevenbf\scriptfont0=\sevenbf
      \scriptscriptfont0=\fivebf
      \textfont1=\elevenmib\scriptfont1=\seveni
      \scriptscriptfont1=\fivei
      \textfont2=\elevenbsy\scriptfont2=\sevensy
      \scriptscriptfont2=\fivesy}%
   \def\scr{\scrfonts
      \global\textfont\scrfam=\elevenscr\fam\scrfam\elevenscr}%
   \tt\ttglue=.5emplus.25emminus.15em
   \normalbaselineskip=13pt
   \setbox\strutbox=\hbox{\vrule height 9pt depth 4pt width 0pt}%
   \normalbaselines\rm\singlespaced}%

%       Switch to 12 point type
\def\twelvepoint{\twelvefonts\ninefonts % load 12pt and 9pt fonts if needed
   \def\rm{\fam0\twelverm}%
   \textfont0=\twelverm\scriptfont0=\ninerm\scriptscriptfont0=\sevenrm
   \textfont1=\twelvei\scriptfont1=\ninei\scriptscriptfont1=\seveni
   \textfont2=\twelvesy\scriptfont2=\ninesy\scriptscriptfont2=\sevensy
   \textfont3=\twelveex\scriptfont3=\twelveex\scriptscriptfont3=\twelveex
   \textfont\itfam=\twelveit\def\it{\fam\itfam\twelveit}%
   \textfont\slfam=\twelvesl\def\sl{\fam\slfam\twelvesl}%
   \textfont\ttfam=\twelvett\def\tt{\fam\ttfam\twelvett}%
   \textfont\bffam=\twelvebf
   \scriptfont\bffam=\ninebf
   \scriptscriptfont\bffam=\sevenbf\def\bf{\fam\bffam\twelvebf}%
   \def\mib{%
      \twelvemibfonts\tenmibfonts
      \textfont0=\twelvebf\scriptfont0=\ninebf
      \scriptscriptfont0=\sevenbf
      \textfont1=\twelvemib\scriptfont1=\ninei
      \scriptscriptfont1=\seveni
      \textfont2=\twelvebsy\scriptfont2=\ninesy
      \scriptscriptfont2=\sevensy}%
   \def\scr{\scrfonts
      \global\textfont\scrfam=\twelvescr\fam\scrfam\twelvescr}%
   \tt\ttglue=.5emplus.25emminus.15em
   \normalbaselineskip=14pt
   \setbox\strutbox=\hbox{\vrule height 10pt depth 4pt width 0pt}%
   \normalbaselines\rm\singlespaced}%

%       Switch to 14 point type
\def\fourteenpoint{\fourteenfonts\twelvefonts % load fonts as needed
   \def\rm{\fam0\fourteenrm}%
   \textfont0=\fourteenrm\scriptfont0=\twelverm\scriptscriptfont0=\tenrm
   \textfont1=\fourteeni\scriptfont1=\twelvei\scriptscriptfont1=\teni
   \textfont2=\fourteensy\scriptfont2=\twelvesy\scriptscriptfont2=\tensy
   \textfont3=\fourteenex\scriptfont3=\fourteenex
      \scriptscriptfont3=\fourteenex
   \textfont\itfam=\fourteenit\def\it{\fam\itfam\fourteenit}%
   \textfont\slfam=\fourteensl\def\sl{\fam\slfam\fourteensl}%
   \textfont\bffam=\fourteenbf
   \scriptfont\bffam=\twelvebf
   \scriptscriptfont\bffam=\tenbf\def\bf{\fam\bffam\fourteenbf}%
   \def\mib{%
      \fourteenmibfonts\twelvemibfonts\tenmibfonts
      \textfont0=\fourteenbf\scriptfont0=\twelvebf
      \scriptscriptfont0=\tenbf
      \textfont1=\fourteenmib\scriptfont1=\twelvemib
      \scriptscriptfont1=\tenmib
      \textfont2=\fourteenbsy\scriptfont2=\tenbsy
      \scriptscriptfont2=\tenbsy}%
   \def\scr{\scrfonts
      \global\textfont\scrfam=\fourteenscr\fam\scrfam\fourteenscr}%
   \normalbaselineskip=17pt
   \setbox\strutbox=\hbox{\vrule height 12pt depth 5pt width 0pt}%
   \normalbaselines\rm\singlespaced}%

%       Switch to 16 point type
\def\sixteenpoint{\sixteenfonts\fourteenfonts\twelvefonts % load fonts as
%%needed
   \def\rm{\fam0\sixteenrm}%
   \textfont0=\sixteenrm\scriptfont0=\fourteenrm\scriptscriptfont0=\twelverm
   \textfont1=\sixteeni\scriptfont1=\fourteeni\scriptscriptfont1=\twelvei
   \textfont2=\sixteensy\scriptfont2=\fourteensy\scriptscriptfont2=\twelvesy
   \textfont3=\sixteenex\scriptfont3=\sixteenex\scriptscriptfont3=\sixteenex
   \textfont\itfam=\sixteenit\def\it{\fam\itfam\sixteenit}%
   \textfont\slfam=\sixteensl\def\sl{\fam\slfam\sixteensl}%
   \textfont\bffam=\sixteenbf%
   \scriptfont\bffam=\fourteenbf%
   \scriptscriptfont\bffam=\twelvebf\def\bf{\fam\bffam\sixteenbf}%
   \def\mib{%
      \sixteenmibfonts\fourteenmibfonts\twelvemibfonts
      \textfont0=\sixteenbf\scriptfont0=\fourteenbf
      \scriptscriptfont0=\twelvebf
      \textfont1=\sixteenmib\scriptfont1=\fourteenmib
      \scriptscriptfont1=\twelvemib
      \textfont2=\sixteenbsy\scriptfont2=\fourteenbsy
      \scriptscriptfont2=\twelvebsy}%
   \def\scr{\scrfonts
      \global\textfont\scrfam=\sixteenscr\fam\scrfam\sixteenscr}%
   \normalbaselineskip=20pt%
   \setbox\strutbox=\hbox{\vrule height 14pt depth 6pt width 0pt}%
   \normalbaselines\rm\singlespaced}%

%       Switch to 20 point type
\def\twentypoint{\twentyfonts\sixteenfonts\fourteenfonts % load fonts
   \def\rm{\fam0\twentyrm}%
   \textfont0=\twentyrm\scriptfont0=\sixteenrm\scriptscriptfont0=\fourteenrm
   \textfont1=\twentyi\scriptfont1=\sixteeni\scriptscriptfont1=\fourteeni
   \textfont2=\twentysy\scriptfont2=\sixteensy\scriptscriptfont2=\fourteensy
   \textfont3=\twentyex\scriptfont3=\twentyex\scriptscriptfont3=\twentyex
   \textfont\itfam=\twentyit\def\it{\fam\itfam\twentyit}%
   \textfont\slfam=\twentysl\def\sl{\fam\slfam\twentysl}%
   \textfont\bffam=\twentybf
   \scriptfont\bffam=\sixteenbf
   \scriptscriptfont\bffam=\fourteenbf\def\bf{\fam\bffam\twentybf}%
   \def\mib{%
      \twentymibfonts\sixteenmibfonts\fourteenmibfonts%
      \textfont0=\twentybf\scriptfont0=\sixteenbf%
      \scriptscriptfont0=\fourteenbf%
      \textfont1=\twentymib\scriptfont1=\sixteenmib%
      \scriptscriptfont1=\fourteenmib%
      \textfont2=\twentybsy\scriptfont2=\sixteenbsy%
      \scriptscriptfont2=\fourteenbsy}%
   \def\scr{\scrfonts
      \global\textfont\scrfam=\twentyscr\fam\scrfam\twentyscr}%
   \normalbaselineskip=24pt
   \setbox\strutbox=\hbox{\vrule height 17pt depth 7pt width 0pt}%
   \normalbaselines\rm\singlespaced}%

%       Switch to 24 point type
\def\twentyfourpoint{\twentyfourfonts\twentyfonts\sixteenfonts % load fonts
   \def\rm{\fam0\twentyfourrm}%
   \textfont0=\twentyfourrm\scriptfont0=\twentyrm\scriptscriptfont0=\sixteenrm
   \textfont1=\twentyfouri\scriptfont1=\twentyi\scriptscriptfont1=\sixteeni
   \textfont2=\twentyfoursy\scriptfont2=\twentysy\scriptscriptfont2=\sixteensy
   \textfont3=\twentyfourex\scriptfont3=\twentyfourex
      \scriptscriptfont3=\twentyfourex
   \textfont\itfam=\twentyfourit\def\it{\fam\itfam\twentyfourit}%
   \textfont\slfam=\twentyfoursl\def\sl{\fam\slfam\twentyfoursl}%
   \textfont\bffam=\twentyfourbf
   \scriptfont\bffam=\twentybf
   \scriptscriptfont\bffam=\sixteenbf\def\bf{\fam\bffam\twentyfourbf}%
   \def\mib{%
      \twentyfourmibfonts\twentymibfonts\sixteenmibfonts%
      \textfont0=\twentyfourbf\scriptfont0=\twentybf
      \scriptscriptfont0=\sixteenbf
      \textfont1=\twentyfourmib\scriptfont1=\twentymib
      \scriptscriptfont1=\sixteenmib
      \textfont2=\twentyfourbsy\scriptfont2=\twentybsy
      \scriptscriptfont2=\sixteenbsy}%
   \def\scr{\scrfonts
      \global\textfont\scrfam=\twentyfourscr\fam\scrfam\twentyfourscr}%
   \normalbaselineskip=28pt
   \setbox\strutbox=\hbox{\vrule height 19pt depth 9pt width 0pt}%
   \normalbaselines\rm\singlespaced}%

%------------------------------------------------------------------------
% Starting defaults:

%>>> EOF TXSfonts.tex <<<

%% file: TXSmacs.tex                             TeXsis version 2.15
%  $Revision: 16.0 $  :  $Date: 1992/08/08 22:24:39 $  :  $Author: myers $
% INTERLINE SPACING.

\def\singlespaced{% sets interline spacing to \normalbaselineskip
   \baselineskip=\normalbaselineskip            % reset interline
}

%>>> EOF TXSmacs.tex <<<

%% TXSenvmt.tex                                 TeXsis version 2.15
%  $Revision: 16.2 $  :  $Date: 1992/09/08 14:46:24 $  :  $Author: myers $
%=======================================================================*
% LIST ENVIRONMENTS:
%
%       All 'list' environments are surrounded by a certain amount of skip.
% These skips are: \EnvTopskip, \EnvBottomskip, \EnvLeftskip, \EnvRightskip
% These are set here, but you may change them if you like.

\newskip\EnvTopskip     \EnvTopskip=\medskipamount      % skip before
\newskip\EnvBottomskip  \EnvBottomskip=\medskipamount   % skip after
\newskip\EnvLeftskip    \EnvLeftskip=2\parindent        % left indent
\newskip\EnvRightskip   \EnvRightskip=\parindent        % right margin in too
\newskip\EnvDelt@skip   \EnvDelt@skip=0pt               % nested skip amount
\newcount\@envDepth     \@envDepth=\z@                  % depth of environments

%       \beginEnv{<name>} does common processing for starting a
% list environment. \endEnv{<name>} does common end procesing,
% and checks the name to make sure that the environments balance.

\def\beginEnv#1{%  begin a ``list'' environment
   \begingroup                          % environment is inside a group
   \def\@envname{#1}%                   % save envmt name, to check at end
   \ifvmode\def\@isVmode{T}%            % remember existing V/H mode
   \else\def\@isVmode{F}\vskip 0pt\fi   % hmode: force vertical mode
   \ifnum\@envDepth=\@ne\parindent=\z@\fi % 1st envmt?  no parindent
   \global\advance\@envDepth by \@ne    % increment level by one
   \EnvDelt@skip=\baselineskip          % \EnvDelt@skip is \baselineskip
   \advance\EnvDelt@skip by-\normalbaselineskip%  minus \normalbaselineskip
   \@setenvmargins\EnvLeftskip\EnvRightskip % now adjust margins.
   \setenvskip{\EnvTopskip}%            % get appropriate topskip
   \vskip\skip@\penalty-500             % and do it (good place to break)
   }

\def\endEnv#1{%         end a ``list'' environment
   \ifnum\@envDepth<1                   % is there nothing open?
      \emsg{> Tried to close ``#1'' environment, but no environment open!}%
      \begingroup                       % \endgroup below would produce error
   \else                                % No: there was an environment open
      \def\test{#1}%                    % was right thing closed?
      \ifx\test\@envname\else           % check that the names match
         \emsg{> Miss-matched environments!}%
         \emsg{> Should be closing ``\@envname'' instead of ``\test''}%
      \fi                               %
   \fi                                  %
   \vskip 0pt                           % force vmode, finish any paragraph
   \setenvskip\EnvBottomskip            % and skip a bottomskip which is
   \vskip\skip@\penalty-500             %    appropriate here (good breakpoint)
   \xdef\@envtemp{\@isVmode}%           % save \@isVmode for outside group
   \endgroup                            % end grouping of environment
   \global\advance\@envDepth by -\@ne   % decrement environment level
   \if F\@envtemp\vskip-\parskip\noindent\fi % no indent if didn't start in
%%vmode
   }

%       \setenvskip chooses a skip amount based on the current \@envDepth,
% and puts it into \skip@, which is a temporary skip register.

\def\setenvskip#1{\skip@=#1 \divide\skip@ by \@envDepth}

%       \@setenvmargins{<left amount>}{<right amount>} adjusts the area of
%  the page to be used by changing \rightskip, \leftskip, and the display
% sizes.  Values given should be skips.

\def\@setenvmargins#1#2{%       set left and right margins
   \advance \leftskip  by #1    \advance \displaywidth by -#1   %
   \advance \rightskip by #2    \advance \displaywidth by -#2   %
   \advance \displayindent by #1}                               %

% \@eatpar gets rid of any \par that follows.  We have to use \futurelet
% to avoid problems when what follows is a \def or somesuch.

\def\@eatpar{\futurelet\next\@testpar}
\def\@testpar{\ifx\next\par\let\@next=\@@eatpar\else\let\@next=\relax\fi\@next}
\long\def\@@eatpar#1{\relax}

%---------------------------------*
%        \TeXexample is an environment for TeX examples. The only special
% characters are <space>, which does the usual thing, and "|", which is the
% escape character. To use a macro in this environment, begin the name with
% "|" instead of "\". In particular, |char`|| gives a |. The enviroment is
% ended with |endTeXexample, NOT \endTeXexample:
%
%       \TeXexample
%           <TeX stuff>
%       |endTeXexample
%
%  <TeX stuff> is printed in \tt type indented by \EnvLeftskip using
%  \obeylines and \obeyspaces and single spaced.  If necessary, it will
%  be split across pages.

\def\TeXexample{\beginEnv{TeXexample}%  % TeX examples
   \vskip\EnvDelt@skip                  % add some extra skip above
   \parskip=\z@ \parindent=\z@          % set \par indentation to zero
   \baselineskip=\normalbaselineskip    % singlespaced
   \def\par{\leavevmode\endgraf}%       % \par also gives \leavevmode
   \obeylines                           % respect line endings
   \catcode`|=\z@                       % make | the escape character
   \ttverbatim                          % begin \tt type in a group
   \@eatpar}%                           % eat initial \par

                 % end the environment

%------------------------------*
%       \ttverbatim makes everything except "|" into \other, then switches
% into \tt type. "|" is made active by \TeXquoteon and is made the escape
% character by \TeXexample and \begintt.

\chardef\other=12

\def\ttverbatim{\begingroup                     % begin a group
   \catcode`\(=\other \catcode`\)=\other        % make everything "other"
   \catcode`\"=\other \catcode`\[=\other        %
   \catcode`\]=\other                           %
   \let\do=\uncatcode \dospecials               %
   \obeyspaces \obeylines                       % obey line ends and spaces
   \def\n{\vskip\baselineskip}%                 % \n gives a new line
   \tt}                                         % switch to typewriter type

\def\uncatcode#1{\catcode`#1=\other}            % make a character "other"

{\obeyspaces\gdef {\ }}                        % space gives \ , not \space

%-----------------------------------*
%       \TeXquoteon makes "|" active and a TeX quote. Anything enclosed
% in | ... | is printed verbatim in \tt type; ^^M's are ignored.
% \TeXquoteoff restores the normal |.

\def\TeXquoteon{\catcode`\|=\active}            % turn on "TeX quotes"
            % turn off "TeX quotes"

{\TeXquoteon\obeylines                          % active "|" calls \ttverbatim
   \gdef|{\ifmmode\vert\else                    % | is \vert in math mode, but
     \ttverbatim \spaceskip=\ttglue             % to use \tt type
     \let^^M=\ %                                % and to ignore ^^M
     \let|=\endgroup                            % next | turns it off
     \fi}                                       % end of \gdef|
}
%       \ttvert| gives a vertical bar in \tt type.  Use anywhere.

%---------------------------------*
%    \raggedcenter centers ragged lines, e.g. for titles.  Each line will
% be as long as possible, centered. Line breaks in the manuscript file are
% ignored.

\def\raggedcenter{%     center lines as long as they can be
    \flushenv                           % do common stuff
    \advance\leftskip\z@ plus4em        % add stretch to sides
    \advance\rightskip\z@ plus 4em      % add stretch to sides
    \spaceskip=.3333em \xspaceskip=.5em %
    \pretolerance=9999 \tolerance=9999  %
    \hyphenpenalty=9999 \exhyphenpenalty=9999   % no hyphens!
    \@eatpar}                           %

\def\endraggedcenter{\endflushenv}              % ends like all flushenv's

\def\flushenv{%  common startup for all flush/center environments
    \vskip \z@                          % force vertical mode
    \bgroup                             % begin grouping
     \def\flushhmode{F}%                % flag: not hmode
     \parindent=\z@  \parfillskip=\z@}  %

\def\endflushenv{% common end to all flush/center environments
   \ifhmode\endgraf\fi                          % if hmode, end \par
   \if T\flushhmode \egroup\hss\fi              % close group and box, or
   \egroup}                                     % end the grouping

% >>> EOF TXSenvmt.tex <<<

% file: TXSsymb.tex                              TeXsis version 2.15
% $Revision: 16.1 $  :  $Date: 1992/08/11 23:41:25 $  :  $Author: myers $
%======================================================================*

%  \slashchar puts a slash through a character to represent contraction
%  with Dirac matrices.

\def\slashchar#1{\setbox0=\hbox{$#1$}           % set a box for #1
   \dimen0=\wd0                                 % and get its size
   \setbox1=\hbox{/} \dimen1=\wd1               % get size of /
   \ifdim\dimen0>\dimen1                        % #1 is bigger
      \rlap{\hbox to \dimen0{\hfil/\hfil}}      % so center / in box
      #1                                        % and print #1
   \else                                        % / is bigger
      \rlap{\hbox to \dimen1{\hfil$#1$\hfil}}   % so center #1
      /                                         % and print /
   \fi}                                         %

%       \simge and \simle make the "greater than about" and the "less
% than about" symbols with spacing as relations.

\def\simge{%  ``greater than about'' symbol
    \mathrel{\rlap{\raise 0.511ex
        \hbox{$>$}}{\lower 0.511ex \hbox{$\sim$}}}}

\def\simle{%  ``less than about'' symbol
    \mathrel{\rlap{\raise 0.511ex
        \hbox{$<$}}{\lower 0.511ex \hbox{$\sim$}}}}

                              % synonym for \simge
                              % synonym for \simle

% ---------- Abbreviations for units

                     % 10^12 electron volts
                     % 10^9  electron volts
                     % 10^6  electron volts
                     % 10^3  electron volts
                       % 1     electron volt

                       % 10^-27 cm^2
                 % 10^-30 cm^2
\def\nb{{\rm nb}}                       % 10^-33 cm^2
                       % 10^-36 cm^2
\def\fb{{\rm fb}}                       % 10^-39 cm^2

         % cm^-2s^-1 for luminosity

%=======================================================================*
%	GEM TDR macros

\newcount\subsubsecno \subsubsecno=0		% subsubsections too
\def\chapsym{}
\def\ch@psymdash{}
\def\ch@psymdot{}

\def\GEMdoc{%
   \openin\lfilein=labeldefs.tmp		% open
   \ifeof\lfilein\closein\lfilein		% if empty, just close
   \else\closein\lfilein			% otherwise close
   \input labeldefs.tmp \relax\fi\writedefs				% for forward refs.
   \twelvepoint\baselineskip=16pt%		% font
   \parskip=\medskipamount%			% roughly like Word
   \GEMsuperrefstrue				% super refs
   \let\newchap=\GEMchapter%			% GEM chapter
   \let\newsec=\GEMsection%			% GEM section
   \let\subsec=\GEMsubsection%			% GEM subsection
   \let\subsubsec=\GEMsubsubsection%		% GEM subsubsection
   \let\subsubsubsec=\GEMsubsubsubsection%	% GEM sub^3section
   \let\eqnres@t=\GEMeqnres@t%			% GEM reset (chapter ID)
   \let\listrefs=\GEMlistrefs			% no \eject, etc.
   \let\ref=\GEMref				% no brackets
   \let\nref=\GEMnref				% ""
   \let\xref=\GEMxref				% ""
   \let\cite=\GEMcite				% super or [] with spacing
   \let\nfig=\GEMnfig				% does a \topinsert
   \footline={\hss\tenrm\ch@psymdash\folio\hss}%% page no with chapter
}

\def\MIB{\mib}

\def\GEMchapter#1#2{%				% GEM chapter
   \gdef\chapsym{#1}%				% chapter number for users
   \gdef\ch@psymdash{#1--}%			% with a --
   \gdef\ch@psymdot{#1.}%			% with a .
   \vfill\supereject				% clear
   \raggedcenter%
      {\fourteenpoint\bf\MIB\ch@psymdot\ \ #2}%	% centered heading
   \endraggedcenter%
   \bigskip\bigskip				% space
   \global\secno=0\global\subsecno=0		% reset
   \global\subsubsecno=0\global\meqno=1		% reset
}

\def\GEMsection#1{%				% GEM section
   \bigbreak\bigskip				% skip
   \global\advance\secno by1%			% new number
   \gdef\secsym{\ch@psymdot\the\secno.}%	% section ident
   \message{(\secsym #1)}%			% announce
   \global\subsecno=0\global\subsubsecno=0	% reset
   \global\meqno=1				% reset
   \GEMhe@ding{\bf\MIB\secsym\ \ }{#1}%		% heading text
   \writetoca{{\secsym}{#1}}%			% add to TOC
}

\def\GEMsubsection#1{%				% GEM subsection
   \bigbreak\bigskip				% skip
   \global\advance\subsecno by1			% new number
   \gdef\subsecsym{\secsym\the\subsecno.}%	% section ident
   \message{(\subsecsym #1)}%			% announce
   \global\subsubsecno=0%			% reset
   \GEMhe@ding{\bf\MIB\subsecsym\ \ }{#1}%	% heading text
   \writetoca{\string\quad {\subsecsym} {#1}}% 	% add to TOC
}

\def\GEMsubsubsection#1{%			% GEM subsubsection
   \bigbreak\bigskip				% skip
   \global\advance\subsubsecno by1		% new number
   \gdef\subsubsecsym{\subsecsym\the\subsubsecno.}% section ident
   \message{(\subsubsecsym #1)}% 		% announce
   \GEMhe@ding{\sl\subsubsecsym\ \ }{#1}% 	% heading text
   \writetoca{\string\quad {\subsubsecsym} {#1}}% write to TOC
}%
\def\GEMsubsubsubsection#1{%			% GEM subsubsubsection
   \bigbreak\bigskip				% skip
   \GEMhe@ding{}{\sl #1}% 			% heading text
}%

\def\GEMhe@ding#1#2{%				% Number #1, text #2
   \vbox{\raggedright\tolerance=10000		% ragged, unbreakable
      \setbox0=\hbox{#1}%			% box = number
      \dimen0=\wd0%				% get width
      \hangindent=\dimen0 \hangafter=1%		% hanging indent
      {\noindent #1#2}%				% text
      \medskip}\nobreak%			% unbreakable skip
}

%	Just like \listrefs, but no \vfill\eject or \centerline
\def\GEMlistrefs{%
   \footatend\relax\immediate\closeout\rfile\writestoppt
   \bigbreak\bigskip
   \vbox{{\bf\noindent References}\medskip}
   \nobreak
   {\frenchspacing\parindent=20pt\escapechar=` \input refs.tmp\vskip0pt}%
   \nonfrenchspacing}

%	Centered table macro. It needs a standard \halign sample line,
% followed by the text:

\def\GEMtable{%
   \hbox to\hsize\bgroup\hss\vbox\bgroup	% for centering
   \def~{\phantom{0}}%				% ~ is phantom digit
   \tabskip=0pt\halign\bgroup}			% start \halign

\def\endGEMtable{%
   \egroup\egroup\hss\egroup}			% end groups

%	References. \nref now defines the control sequence to be just
% the number; \cite makes a citation with its argument being the
% references, e.g. \cite{\a} or \cite{\a,\b}.

%	Insert citation mark. Note there is no lookahead!

\newif\ifGEMsuperrefs	\GEMsuperrefstrue

\def\GEMcite#1{%
   \begingroup
      \let\@sf=\empty				% initialize to none
      \ifhmode\edef\@sf{%                       % save spacefactor
         \spacefactor\the\spacefactor}\/\fi     %
      \ifGEMsuperrefs                           % superscript references?
         $\relax{}^{#1}$\@sf%                   % superscript
      \else {}~[{#1}]\@sf\fi%                   % [] style
   \endgroup}

\def\GEMref{\GEMcite{\the\refno}\GEMnref}

%	No brackets in definition, and a . in the printout -- otherwise
% Ginsparg's \nref.
\def\GEMnref#1{\xdef#1{\the\refno}\writedef{#1\leftbracket#1}%
\ifnum\refno=1\immediate\openout\rfile=refs.tmp\fi
\global\advance\refno by1\chardef\wfile=\rfile\immediate
\write\rfile{\noexpand\item{#1.\ }\reflabeL{#1\hskip.31in}\pctsign}\findarg}

%	\xref is just a dummy -- already stripped.
\def\GEMxref#1{#1}

%	Figures are done as \topinsert's. #1 is the figure name, #2
% is the space to reserve for it, and #3 is the caption. (This may
% break with a very long caption, in which case do it by hand!

\def\GEMnfig#1#2#3{%
   \xdef#1{\ch@psymdash\the\figno}%		% save counter
   \writedef{#1\leftbracket#1}%			% and save it
   \topinsert%					% insert
      #2%					% vmode command
      \bigskip%					% skip down
      \GEMcaption{Fig.~#1}{#3}%			% caption
   \endinsert%					% end
   \global\advance\figno by1}			% advance counter

\newbox\c@pbox
\newcount\c@plines

%	\GEMcaption makes a caption. Note that it includes no spacing;
% you must add spacing above or below by hand.
\def\GEMcaption#1#2{%				% caption
   \begingroup\tenpoint				% all in 10pt
   \global\setbox\c@pbox=\vbox{%		%
      \setbox0=\hbox{\noindent #1:\ \ }%	% set a box
      \hangindent=\wd0 \hangafter=1		% and set hanging indent
      \noindent\hbox{\noindent #1:\ \ }{#2}%	% get text
      \vskip0pt\relax				% force vmode
      \global\c@plines=\prevgraf}%		% lines in caption
   \ifnum\@ne=\c@plines%			% if 1 line
      \global\setbox\c@pbox=\vbox{%		% set box
      \noindent\hfil				% centered
      \hbox{\noindent#1:\ \ }{#2}\hfil}\fi%	% text for 1 line
   \centerline{\box\c@pbox}%			% center box
   \endgroup					% end 10pt
}

%	\GEMepsf#1#2 inserts a EPS figure for Tomas Rokicki's dvips.
% #1 is the file name, and #2 is the vertical size. If it is not
% available, a vertical skip is inserted.

\def\GEMepsf#1#2{%
   \ifx\undefined\epsfbox			% if epsf not loaded
      \vskip#2\relax				% just leave space
   \else					% if loaded
      \epsfysize=#2				% scale to size
      \epsfbox{#1}%				% insert figure
   \fi
}

\catcode`@=12

%======================================================================*
%	Choose format
\def\bigans{b }
\def\littleans{l }

\everyjob={%
   \message{ big or little (b/l)? }\read-1 to\answ
   \ifx\answ\bigans
      \message{This will come out unreduced.}
      \PPTbig
   \else
      \if\answ\littleans
         \message{This will come out reduced.}
         \PPTlittle
      \else
         \message{No format defined}
      \fi
   \fi}

%	Execute \everyjob if gemmac is read in. Use x to select no
% format for gemmac.fmt.
\message{ big or little (b/l)? }\read-1 to\answ
\ifx\answ\bigans
   \message{This will come out unreduced.}
   \PPTbig
\else
   \if\answ\littleans
      \message{This will come out reduced.}
      \PPTlittle
   \else
      \message{No format defined}
   \fi
\fi

%
%
%
%
%	GEMMAC_SYM.TEX
%
%Curly letters
%

\def\CD{{\cal D}}

\def\CG{{\cal G}}

\def\CL{{\cal L}}

\def\CS{{\cal S}}

\def\CU{{\cal U}}

%
% TECHNICOLOR SYMBOL MACROS
%
\def\ts{\thinspace}
\def\ra{\rightarrow}
\def\ol{\overline}

\def\atc{\alpha_{TC}}

\def\acrit{\alpha_C}

\def\gmm{\gamma_m}

\def\Getc{G_{ETC}}

\def\Few{F_\pi}

\def\W+-{W^\pm}
\def\Z0{Z^0}

\def\condt{\langle \overline T T\rangle}

\def\condtc{{\langle \ol T T \rangle}_{TC}}
\def\condetc{{\langle \ol T T \rangle}_{ETC}}

\def\Ms+-{M^2_\pm}

\def\M+-{M_\pm}

\def\suc{SU(3)}

\def\tro{\rho_T}

\def\tpi{\pi_T}
\def\tpiqq{\pi_{\ol Q Q}}

\def\tpilq{\pi_{\ol L Q}}

\def\tpiql{\pi_{\ol Q L}}

\def\kslash{\raise.15ex\hbox{/}\kern-.57em k}
\def\pslash{\raise.15ex\hbox{/}\kern-.57em p}
%
%
%   Phenomenology macros
%
%	Symbols, etc.
\def\simge{\mathrel{%
   \rlap{\raise 0.511ex \hbox{$>$}}{\lower 0.511ex \hbox{$\sim$}}}}
\def\simle{\mathrel{
   \rlap{\raise 0.511ex \hbox{$<$}}{\lower 0.511ex \hbox{$\sim$}}}}
% \slashchar puts a slash through a character to represent contraction
% with Dirac matrices.
\def\slashchar#1{\setbox0=\hbox{$#1$}           % set a box for #1
   \dimen0=\wd0                                 % and get its size
   \setbox1=\hbox{/} \dimen1=\wd1               % get size of /
   \ifdim\dimen0>\dimen1                        % #1 is bigger
      \rlap{\hbox to \dimen0{\hfil/\hfil}}      % so center / in box
      #1                                        % and print #1
   \else                                        % / is bigger
      \rlap{\hbox to \dimen1{\hfil$#1$\hfil}}   % so center #1
      /                                         % and print /
   \fi}                                         %

                                    % right-hand spacing
%%%%%%%%%%%%%%%%%%%%%%%%%%%%%%%%%%%%%%%%%%%%%%%%%%%%%%%%%%%%%%

%       Symbol definitions
%

%\def\etcell{E_{{T,\,\rm cell}}}

\def\CG{{\cal G}}
\def\CL{{\cal L}}

\def\cm{{\rm cm}}

\def\fb{{\rm fb}}
\def\gev{{\rm GeV}}
\def\h{H^0}

\def\nb{{\rm nb}}
\def\ol{\overline}
\def\ub{\underbar}

\def\ra{\rightarrow}
\def\ecm{\sqrt{s}}

\def\tev{{\rm TeV}}

\def\tpi{\pi_T}
\def\uhly{10^{41} \ts {\rm cm}^{-2}}

\def\uhl{10^{34} \ts {\rm cm}^{-2} \ts {\rm s}^{-1}}

\def\half{{1 \over {2}}}
\def\thalf{\textstyle{{1 \over {2}}}}
\def\fourth{{1 \over {4}}}
\def\tfourth{\textstyle{{1 \over {4}}}}

\overfullrule=0pt

\ifx\undefined\GEMdoc
\fi

\GEMdoc
\TeXquoteon					% make | a TeX quote
%---------------------------------------
%	The following lines should all be commented out for single
% column format:
% \input TXSdcol
%\setdoublecolumns{6.50in}{9.00in}{3.00in}	% double column format
%\tenpoint\baselineskip=13pt			% 10pt
%\elevenpoint\baselineskip=14pt			% 11pt
%\rightskip=0pt plus 2em\relax			% ragged right
%---------------------------------------
%	The following lines should all be commented out for double
% column format:
\def\doublecolumns{\relax}
\def\enddoublecolumns{\relax}

%---------------------------------------

\def\bull{$\bullet$}

\def\half{{\textstyle{1\over 2}}} %puts a small half in a displayed eqn
\def\fourth{{\textstyle{1\over 4}}}
\def\ub{\underbar}

\def\mev{\rm {MeV}}
\def\Metc{M_{ETC}}
%%%%%%%%%%%%%%%%%%%%%%%%%%%%%%%%%%%%%%%%%%%%%%%%%%%%%%%%%%%%%%%%%%%%%%%%%%%

%\draft

%------------------------------------------------------------------------
%\headline={\ifnum\pageno=1\firstheadline\else
%\ifodd\pageno\rightheadline \else\leftheadline\fi\fi}
%\def\firstheadline{\hfil}
%\def\rightheadline{\hfil}
%\def\leftheadline{\hfil}
%	\footline={\ifnum\pageno=1\firstfootline\else\otherfootline\fi}
%\def\firstfootline{\rm\hss\folio\hss}
%\def\otherfootline{\hfil}
%
%\font\twelvebf=cmbx10 scaled\magstep 1
%\font\twelverm=cmr10 scaled\magstep 1
%\font\twelveit=cmti10 scaled\magstep 1
%\font\elevenbfit=cmbxti10 scaled\magstephalf
%\font\elevenbf=cmbx10 scaled\magstephalf
%\font\elevenrm=cmr10 scaled\magstephalf
%\font\elevenit=cmti10 scaled\magstephalf
%\font\bfit=cmbxti10
%\font\tenbf=cmbx10
%\font\tenrm=cmr10
%\font\tenit=cmti10
%\font\ninebf=cmbx9
%\font\ninerm=cmr9
%\font\nineit=cmti9
%\font\eightbf=cmbx8
%\font\eightrm=cmr8
%\font\eightit=cmti8
%\parindent=1.5pc
\hsize=6.0truein
\vsize=8.5truein
%\nopagenumbers
\def\myfoot#1#2{{\baselineskip14.4pt plus 0.3pt\footnote{#1}{#2}}}
\doublecolumns
%\magnification=1200
%\hskip4.6in\vbox{\baselineskip12pt\hbox{BUHEP--94--2}}
%\bigskip\bigskip
{\Title{\vbox{\baselineskip12pt\hbox{BUHEP--94--2}}}
{AN INTRODUCTION TO TECHNICOLOR\myfoot{$^{\dag}$}{Lectures given
June~30--July~2, 1993 at the Theoretical Advanced Studies Institute,
\hfil\break University of Colorado, Boulder.}}}

%\baselineskip=16pt
%\vglue 0.8cm
\centerline{Kenneth  Lane\myfoot{$^{\ddag }$}{lane@buphyc.bu.edu}}
\smallskip
\centerline{Department of Physics, Boston University}
\smallskip
\centerline{590 Commonwealth Avenue, Boston, MA 02215}
%\baselineskip=12pt
%\centerline{Boston, MA 02215, USA}
%\vglue 0.8cm
\vskip.3in
\centerline{\bf Abstract}
%\vglue 0.3cm
%{\rightskip=3pc
% \leftskip=3pc
%{\baselineskip=12pt\noindent
In these lectures we present the motivation for dynamical electroweak
symmetry breaking and its most popular realization, technicolor. We
introduce the basic ideas of technicolor and its companion theory of
flavor, extended technicolor. We review the classical theory of
technicolor, based on naive scaling from quantum chromodynamics, and
discuss the classical theory's fatal flaws. Finally, we describe the
principal attempt to correct these flaws, the theory of walking
technicolor.
%\vglue 0.6cm}}

\bigskip
%\Date{1/94}
\vfil\eject
%\twelverm
%\baselineskip=14pt

\newsec{WHY TECHNICOLOR?}

In the first part of these lectures we describe the motivations and virtues
of technicolor and extended technicolor---dynamical theories of
electroweak and flavor symmetry. We then give an overview of the
``classical'' theory of technicolor, using arguments based on scaling from
QCD. We discuss the theoretical and phenomenological problems of
technicolor and extended technicolor and summarize the main attempts to
overcome them.

\medskip
\noindent{\it 1.1 The Importance of Electroweak Symmetry Breaking}
\smallskip

	The theoretical elements of the standard $\suc\otimes SU(2)\otimes
U(1)$ gauge model of strong and electroweak interactions have been in place
for more than 20~years.%
\nref\tdr{Much of the discussion in this section appeared in a similar form
in K.~Lane, {\it The Next Collider}, invited talk in the Proceedings of the
Conference on High Energy Physics with Colliding Beams, Yale University,
October 2--3, 1992 and in the Preface of the {\it GEM Technical Design
Report}; GEM TN-93-262, SSCL-SR-1219; Submitted by the GEM Collaboration to
the Superconducting Super Collider Laboratory (April 30, 1993)}%
\nref\sm{S.~L.~Glashow, {\it Nucl.~Phys.~}~{\bf 22} (1961) 579\semi
S.~Weinberg, {\it Phys.~Rev.~Lett.}~{\bf 19} (1967) 1264 \semi
A.~Salam, in Proceedings of the 8th Nobel Symposium on Elementary Particle
Theory, Relativistic Groups and Analyticity, edited by N.~Svartholm
(Almquist and Wiksells, Stockholm, 1968), p.~367\semi
H.~Fritzsch, M.~Gell--Mann, H.~Leutwyler, {\it Phys.~Lett.}~{\bf 47B}
(1973) 365\semi
D.~Gross and F.~Wilczek, {\it Phys.~Rev.~Lett.}~{\bf 30} (1973) 1343\semi
H.~D.~Politzer, {\it Phys.~Rev.~Lett.}~{\bf 30} (1973) 1346.}%
\nref\particles{R.~Cahn and G.~Goldhaber, {\it The Experimental
Foundations of Particle Physics}, (Cambridge University Press, 1989)}%
\cite{\tdr,\sm,\particles} In all this time, the standard model has withstood
extremely stringent experimental tests.%
\ref\smtests{For a recent review, see P.~Langacker, M.--X.~Luo and
A.~K.~Mann, {\it Rev.~Mod.~Phys.} {\bf 64} (1992) 87.}
Down to distances of at least $10^{-16}\,\cm$, the basic constituents of
matter are known to be spin--$\thalf$ quarks and leptons. These interact via
the exchange of spin--one gauge bosons: the massless gluons of QCD and the
massless photon and massive $W^\pm$ and $Z^0$ bosons of electroweak
interactions. There are six {\it flavors} each of quarks and
leptons---identical except for mass, charge and color---grouped into three
generations. All the fermions have been found except for the top quark and
the tau neutrino.%
\nref\topmass{According to Ref.~\PDG, the 95\% confidence--level limit on
the top--quark mass, assuming it decays in the standard way, $t \ra W^+ b$,
is $m_t > 91\,\gev$. The limit obtained without assuming dominance of the
standard decay mode is $m_t > 55\,\gev$. The D0 Collaboration at the
Fermilab Tevatron Collider have just reported a new 95\% limit on the mass
of a standard top--quark of $131\,\gev$ ({\it Search for the Top Quark in
$p \ol p$ Collisions at $\ecm = 1.8\,\tev$}, submitted to Physical Review
Letters). While the tau--neutrino has not yet been directly observed, there
is little doubt that it exists and that it and the $\tau^-$ form a standard
lepton doublet. The 95\% limit on its mass is $m_{\nu_\tau} < 35\,\mev$.
See Ref.~\PDG.}%
\nref\PDG{Particle Data Group, K.~Hikasa et al., {\it Phys.~Rev.}~{\bf D45}, S1
(1992).}%
\cite{\topmass,\PDG} If the number of quark--lepton generations is equal to
the number $N_\nu$ of light neutrinos, then there are no more than these
three. The evidence for this comes from precision measurements of the $Z^0$
width at LEP, which give $N_\nu = 2.99 \pm 0.04$ in the standard
model.\cite{\PDG}

The fact that the QCD-color gauge symmetry is exact---in both the
Lagrangian and the ground state of the theory---implies that quarks and
gluons are confined at large distances into color--singlet hadrons and that
they are almost noninteracting at small distances. However, confinement and
asymptotic freedom are not the only dynamics open to gauge theories. Even
though gauge bosons necessarily appear in the Lagrangian without mass,
interactions can make them heavy. This is what happens to the $W^\pm$ and
$Z^0$ bosons: electroweak gauge symmetry is spontaneously broken in the
ground state of the theory, a phenomenon known as the ``Higgs mechanism''.%
\ref\higgs{P.~W.~Anderson, {\it Phys.~Rev.}~{\bf 110} (1958) 827; {\it
ibid.}, {\bf 130} (1963) 439\semi
Y.~Nambu, {\it Phys.~Rev.}~{\bf 117} (1959) 648\semi
J.~Schwinger, {\it Phys.~Rev.}~{\bf 125} (1962) 397\semi
P.~Higgs, {\it Phys.~Rev.~Lett.}~{\bf 12} (1964) 132\semi
F.~Englert and R.~Brout, {\it Phys.~Rev.~Lett.}~{\bf 13} (1964) 321\semi
G.~S.~Guralnik, C.~R.~Hagen and T.~W.~B.~Kibble, {\it Phys.~Rev.~Lett.}~{\bf
13} (1964) 585.}
Finally, fermions in the standard model also must start out massless.
To make quarks and leptons massive, new forces beyond the $SU(3)
\otimes SU(2) \otimes U(1)$ gauge interactions are required. These
additional interactions explicitly break the fermions' flavor symmetry
and communicate electroweak symmetry breaking to them.

	Despite this great body of knowledge, the interactions underlying
electroweak and flavor symmetry breakdowns remain {\it unknown}. The most
important element still missing from this description of particle
interactions is directly connected to electroweak symmetry breaking.  This
may manifest itself as a single new particle---the ``Higgs boson''; it
may be several such bosons; or a replication of all the known particles; or
an infinite tower of new resonances; or something still unimagined. It is
also unknown whether the new interactions required for flavor symmetry
breaking need additional new particles for their implementation. Until the
new dynamics are known, it seems impossible to make further progress in
understanding elementary particle physics.

One very important aspect of electroweak symmetry breaking is known: its
characteristic energy scale of $1\,\tev$. This scale is set by the decay
constant of the three Goldstone bosons transformed via the Higgs mechanism
into the longitudinal components, $W^\pm_L$ and $Z^0_L$, of the weak gauge
bosons:
\eqn\weakscale{ \Few \equiv 2^{-\tfourth} G_F^{-\thalf} =
246\,\gev \ts .}
{\ub {\it New physics must occur near this energy scale.}} New particles
produced in parton scattering processes at this energy may appear as fairly
distinct resonances in weak gauge boson or fermion--antifermion final
states, or only as relatively featureless enhancements of $W_L$ and $Z_L$
boson production or of missing energy. Whatever form the new physics takes,
it was the energy scale of $1\,\tev$\ and the size of typical QCD and
electroweak cross sections at this energy, $\sigma \simeq 1\,\nb$ --
$1\,\fb$, that determined the energy and luminosity requirements of the
Superconducting Super Collider: $\ecm = 40\,\tev$ and $\CL = 10^{33}
-\uhl$.%
\ref\ehlq{E.~Eichten, I.~Hinchliffe, K.~Lane and C.~Quigg,
{\it Rev.~Mod.~Phys.}~{\bf 56} (1984) 579.}

The energy scale of flavor symmetry breaking is {\it not} known. It may lie
anywhere from just above the weak scale, $1\,\tev$, up to the Planck scale,
$M_P\simeq 10^{16}\,\tev$. It is possible that high--energy collisions at
the SSC would have shed light on the flavor problem, but there was no
guarantee. We shall see that technicolor---the most studied theory of
dynamical electroweak and flavor symmetry---provides many signatures of
flavor physics in the TeV~energy range. Their production cross sections
also range from quite large ($\sim 1-10\,\nb$) to very small ($\sim
1-10\,\fb$) at SSC energies and would have been accessible there. The
opportunities are fewer at the LHC, but they are not negligible.

Several scenarios have been proposed for electroweak and flavor symmetries,
and their breaking:\cite{\ehlq}

\medskip

\item{$\bullet$} Standard Higgs models, containing one or more elementary
Higgs boson multiplets. These are generally complex weak doublets. The
minimal model has one doublet and, after symmetry breaking, a single
neutral boson, $\h$, remains after the Higgs mechanism. If Higgs bosons
exist as discernible states, theoretical consistency demands that they lie
below about 700--800~GeV.

\item{$\bullet$} Supersymmetry. The most studied example is the minimal
supersymmetric standard model. In this model there are two Higgs doublets,
and every known particle has a superpartner. It is expected that all the
new particles of the MSSM lie below 1~TeV.

\item{$\bullet$} Models of dynamical electroweak and flavor symmetry
breaking. The most studied proposal is
technicolor--plus--extended--technicolor, with one doublet or one family of
technifermions. In the minimal one--doublet model, the observable
technihadrons are expected near 1.5--2.0~TeV and would require a machine
with very high energy and luminosity (such as the SSC) for their discovery.
More complicated examples, such as the one--family model, may well be
testable with lower energy and luminosity.

\item{$\bullet$} Composite models, in which quarks and leptons are built
of more fundamental constituents. All that we know about the scale of
quark--lepton substructure is that it is greater than 1--2~TeV.\cite{\PDG}
One wants the largest possible energy and highest usable luminosity to
search for substructure. (See the GEM TDR in Ref.~\tdr.)

\medskip

All of these scenarios have certain attractive features. However, as we
shall see in these lectures, they also have undesirable ones. Despite their
apparent problems, the standard Higgs boson, $\h$, charged Higgses,
$H^\pm$, and the supersymmetric partners of all the known particles may
exist and must be sought. The same applies to the dynamical technicolor
scenario described in the rest of these lecture notes.

The difficulties outlined below in Section 1.2 have led to the widespread
belief that none of the familiar descriptions of electroweak and flavor
symmetry breaking is entirely correct. That, in fact, was the most exciting
aspect of SSC physics. We know that there is new physics in the TeV~energy
regime. We do not know exactly what form it will take. But we knew that the
SSC could have reached it. Thus, the termination of the SSC project by the
U.~S.~Congress is an enormous blow to particle physics. Time will tell
whether the blow is fatal. In the meantime, we must pursue the secrets of
electroweak and flavor symmetry breaking as best we can. The best hope for
this in the foreseeable future---the next 10--20~years---is experimentation
at the Large Hadron Collider (LHC), proposed to be built at CERN in the LEP
tunnel. The LHC is less powerful (and less expensive) than the SSC. Yet,
Nature may be kind enough to put electroweak breaking physics, and even
flavor physics, within its reach. Thus, the LHC deserves the full support
of all particle physicists, especially those in the United States who
worked so hard on the physics of electroweak symmetry breaking and on the
SSC.

\vfil\eject
%\medskip
\noindent{\it 1.2 Problems With Elementary Higgs Bosons}
\smallskip

In nonsupersymmetric elementary Higgs boson models, there is no explanation
of why electroweak symmetry breaking occurs and why it has the scale
$\Few$. The Higgs doublet self--interaction potential is $V(\phi) =
\lambda\ts (\phi^\dagger \phi - v^2)^2$, where $v$ is the vacuum expectation
of the Higgs field $\phi$ {\it when} $v^2 \ge 0$. But what dynamics makes
$v^2 > 0$? Where does the value $v \sim \Few = 246\,\gev$ come from?

Elementary Higgs boson models are unnatural. The Higgs boson's mass, $M_H =
\sqrt{2 \lambda} v$, and the vacuum expectation value itself are {\it
quadratically} unstable against radiative corrections. Thus, there is no
natural reason why these two parameters should be much less than the
energy scale at which the essential physics of the model changes, e.g., a
unification scale or the Planck scale.%
\ref\natural{K.~G.~Wilson, unpublished; quoted in L.~Susskind,
{\it Phys.~Rev.}~{\bf D20} (1979) 2619\semi
G.~'t~Hooft, in {\it Recent Developments in Gauge Theories}, edited by
G.~'t~Hooft, et al. (Plenum, New York, 1980).}

A further problem of elementary Higgs boson models is that they are
``trivial''.%
\ref\trivial{See, for
example, R. Dashen and H. Neuberger,
{\it Phys.~Rev.~Lett.}~{\bf 50} (1983) 1897\semi J.~Kuti, L.~Lin and Y.~Shen,
{\it Phys.~Rev.~Lett.}~{\bf 61} (1988) 678\semi A.~Hasenfratz, et
al.~{\it Phys.~Lett.}~{\bf 199B} (1987) 531\semi G.~Bhanot and K.~Bitar,
{\it Phys.~Rev.~Lett.}~{\bf 61} (1988) 798.}
To a good approximation, the self--coupling $\lambda(M)$ of the minimal
one--doublet Higgs boson at the energy scale $M$ is given by
\eqn\lamtriv{
\lambda(M) \cong {\lambda(\Lambda) \over {1 + (24 /16 \pi^2)\ts
\lambda(\Lambda) \ts \log (\Lambda /M)}} \ts.}
This vanishes for all $M$ as the cutoff $\Lambda$
is taken to infinity, hence the
description ``trivial''. This feature has been shown to be true in a
general class of two--Higgs doublet models,%
\ref\dk{R.~S.~Chivukula and D.~Kominis, {\it Phys.~Lett.}~{\bf 304B}
(1993) 152.}
and it is probably true of all Higgs models. Triviality means that
elementary--Higgs Lagrangians must be considered to describe
effective theories. They are meaningful only for scales $M$ below some
cutoff $\Lambda_\infty$ at which new physics sets in. The larger the
Higgs couplings are, the lower the scale $\Lambda_\infty$. This
relationship translates into the so--called triviality bounds
on Higgs masses. For the minimal model, the connection between $M_H$ and
$\Lambda_\infty$ is
\eqn\triv{
M_H(\Lambda_\infty) \cong \sqrt{2 \lambda(M_H)} \ts v = {2 \pi v \over
{\sqrt{3 \log (\Lambda_\infty/M_H)}}} \ts.}
Clearly, the Higgs mass has to be somewhat less than the cutoff in order
for the effective theory to have some range of validity. From lattice--based
arguments,\cite{\trivial} $\Lambda_\infty \simge 2 \pi M_H$. Since $v$ is
fixed at 246~GeV in the minimal model, this implies the triviality bound
$M_H \simle 700\,\gev$. If the standard Higgs boson were to be found with a
mass this large or larger, we would know for sure that additional new
physics is lurking in the range of a few~TeV.

Finally, elementary Higgs models provide no clue to the meaning of flavor
symmetry and the origin of its breaking. The flavor--symmetry breaking
Yukawa couplings of the Higgs boson to fermions are arbitrary free
parameters. As far as we know, this is a logically tenable state of
affairs---that we may not understand flavor until we understand the physics
of the Planck scale---but it is a difficult pill to swallow.

The radiative instability of elementary Higgs boson models may be cured by
supersymmetry.%
\nref\SUSYref{S.~Dimopoulos and H.~Georgi, {\it Nucl.~Phys.~}~{\bf B193}
(1981) 153\semi A.H.~Chamseddine, R.~Arnowitt and P.~Nath,
{\it Phys.~Rev.~Lett.}~{\bf 49} (1982) 970 \semi L.~J.~Hall, J.~Lykken and
S.~Weinberg, {\it Phys.~Rev.}~{\bf D27} (1983) 2359.}%
\nref\HaberKane{For reviews of supersymmetry and its phenomenology, see
H.~E.~Haber and G.~L.~Kane, {\it Phys.~Rept.}~{\bf 117} (1985) 75\semi
S.~Dawson, E.~Eichten and C.~Quigg, {\it Phys.~Rev.}~{\bf D31}(1985) 1581.}%
\cite{\SUSYref, \HaberKane}  In the most popular scenario, supersymmetry is
broken at a very high energy scale. This occurs in a sector of the theory
that communicates only through highly suppressed interactions with the
standard--model sector (including superpartners). Thus, standard--model
interactions {\it look} supersymmetric down to a low energy energy scale
where soft supersymmetry breaking effects become important. This protects
the Higgs bosons' masses, the Higgs vacuum expectation values, and the
masses of superpartners---gluinos, squarks, etc. Furthermore, it offers a
plausible explanation of why the effective supersymmetry breaking, and
electroweak breaking, scale is much less than the Planck mass, $M_P$.
Triviality is not a pressing issue in the minimal supersymmetric
standard model because the Higgs masses are relatively low and, so, the
cutoff $\Lambda_\infty$ may be very high indeed.

Like the ordinary Higgs models, however, supersymmetry makes no attempt to
explain the meaning of flavor symmetry and the origin of its breakdown.
Again, flavor is broken by arbitrary Yukawa couplings put in by hand.
Unlike the ordinary models, supersymmetric Higgs models suffer from a
problem commonly, but incorrectly, assumed to afflict only technicolor:
unacceptably large flavor--changing neutral currents (FCNC).%
\ref\Haber{See H.~E.~Haber, U.~C.~Santa Cruz preprint SCIPP~93/22 (July,
1993) for a good summary of the positive and
negative features of supersymmetry at the weak scale.}
These occur via squark exchange {\it unless} the squark mass matrices are
simultaneously diagonalizable with the quark mass matrices and
the squark masses chosen to be nearly degenerate. This
can be done, but it is not always natural to do so.

\medskip
\noindent{\it 1.3 Eliminating Elementary Higgs Bosons}
\smallskip

To break electroweak symmetry, there is no need for elementary Higgs
bosons. Suppose the world were described by the standard $SU(3) \otimes
SU(2) \otimes U(1)$ Lagrangian with gauge bosons, $n_G = 3$ generations of
quarks and leptons, and nothing else:
\eqn\stdlag{\eqalign{
\CL_{\rm STD} &= -\fourth G_{\mu\nu}^A G^{A\ts \mu \nu}
-\fourth W_{\mu\nu}^a W^{a\ts \mu \nu}
-\fourth W_{\mu\nu}^0 W^{0\ts \mu \nu} \cr
&\quad + \sum_{i = 1}^{n_G} \biggl(\ol q_{i \alpha L} \ts i \gamma_\mu
\CD^\mu_{\alpha \beta} q_{i \beta L} + \ol u_{i \alpha R} \ts i \gamma_\mu
\CD^\mu_{\alpha \beta} u_{i \beta R} + \ol d_{i \alpha R} \ts i \gamma_\mu
\CD^\mu_{\alpha \beta} d_{i \beta R} \cr &\qquad + \ol L_{i L} \ts i \gamma_\mu
\CD^\mu L_{i L} + \ol \ell_{i R} \ts i \gamma_\mu
\CD^\mu \ell_{i R} \biggr) \ts .\cr}}
Here, $SU(3)$--colors for the quarks are labeled by $\alpha = 1,2,3$.
The QCD interaction is mediated by eight gluons, $g^{A = 1,...,8}$. The
electroweak gauge bosons are $W^{a=1,2,3}$ for $SU(2)_{\rm EW}$ and $W^0$
for $U(1)_{\rm EW}$.  The $SU(3)$, $SU(2)$ and $U(1)$ gauge couplings are
$g_s$, $g$ and $g'$ respectively; the field strength for, say, the gluon is
$G_{\mu\nu}^A = \partial_\mu g_\nu^A -\partial_\nu g_\mu^A + g_s f_{ABC}
g_\mu^B g_\nu^C$; the covariant derivative for an electroweak doublet of
quarks, $q_{i \alpha L}$, is
$\CD^\mu_{\alpha \beta} = (\partial^\mu - i g/2\ts \tau_a \ts W^{a\mu} - i
g'/6 \ts W^{0 \mu})\delta_{\alpha \beta} - i g_s \ts (\lambda_A /2)_{\alpha
\beta} \ts g^{A \mu}$; and so on. Note that the chiral nature of quark
and lepton transformation laws under the electroweak gauge group
forbid bare mass terms for these fermions.

Ignore the small electroweak couplings of quarks for now. Then
their interactions respect a large global chiral flavor symmetry,%
\ref\foota{The chiral flavor group includes a vectorial $U(1)$ which is
essentially baryon number. It does not include the corresponding
axial--vectorial $U(1)$ symmetry, which is strongly broken by instanton
effects. See G.~'t~Hooft, {\it Phys.~Rev.~}~{\bf D14} (1976) 3432.}
\eqn\gchi{\CG_{\chi} = SU(2 n_G)_L \otimes SU(2 n_G)_R \ts.}
The strong QCD interactions of quarks cause this chiral symmetry to be
spontaneously broken%
\ref\vector{G.~'t~Hooft in Ref.~\natural\semi
S.~Coleman and E.~Witten, {\it Phys.~Rev.~Lett.}~{\bf 45} (1980) 100\semi
C.~Vafa and E.~Witten, {\it Nucl.~Phys.}~{\bf B234} (1984) 173.}
by the {\it condensates}
\eqn\qcond{
\langle \Omega \vert \ol u_{i \alpha L} u_{j \beta R} \vert \Omega \rangle =
\langle \Omega \vert \ol d_{i \alpha L} d_{j \beta R} \vert \Omega \rangle =
- \delta_{ij} \delta_{\alpha \beta} \Delta_q \ts .}
In Eq.~\qcond, $\vert \Omega \rangle$ is the ground state of QCD whose
symmetry group is $SU(2 n_G)_V$, the diagonal subgroup of $SU(2 n_G)_L
\otimes SU(2 n_G)_R$. Consequently, there are $4 n_G^2-1$ massless
pseudoscalar mesons, commonly called Goldstone bosons, coupling to the
appropriately--defined axial--vector currents with  strength $f_\pi =
93\,\mev$.%
\nref\nambu{Y.~Nambu and G.~Jona--Lasinio, {\it Phys.~Rev.}~{\bf 122}
(1961) 345.}%
\nref\goldstone{J.~Goldstone, {\it Nuovo Cimento}~{\bf 19A} (1961) 154\semi
J.~Goldstone, A.~Salam and S.~Weinberg, {\it Phys.~Rev.}~{\bf 127}
(1962) 965.}%
\cite{\nambu,\goldstone}
\nref\delqcd{A.~Manohar and H.~Georgi, {\it Nucl.~Phys.~}~{\bf B234}
(1984) 189\semi H.~Georgi and L.~Randall,
{\it Nucl.~Phys.~}~{\bf B276} (1986) 241.}
According to Ref.~\delqcd, the quark condensate is approximated by
$\Delta_q \simeq 4 \pi f^3_\pi$.

Now restore the electroweak interactions. The quark parts of the $SU(2)
\otimes U(1)$ currents couple to a normalized linear combination of these
Goldstone bosons with strength $\sqrt{n_G} f_\pi$. These massless states
appear as poles in the polarization tensors, $\Pi_{\mu \nu}^{ab}(q)$, of
the electroweak gauge bosons. Near $q^2 = 0$,
\eqn\pipole{\Pi_{\mu \nu}^{ab}(q) = (q_\mu q_\nu - q^2 g_{\mu \nu})
\left({g_a g_b n_G f_\pi^2 \over {4 \ts q^2}}\right) \ts\ts + \ts\ts
{\rm nonpole \ts\ts terms} \ts.}
Here $a,b = 0,1,2,3$; $g_0 = g'$ and $g_{1,2,3} = -g$. The electroweak
symmetry $SU(2) \otimes U(1)$ has broken down to $U(1)_{\rm EM}$ and the
bosons $W^\pm = (W^1 \mp i W^2)/\sqrt{2}$ and $Z = (g W^3 - g'
W^0)/\sqrt{g^2 + g'^2}$ have acquired mass
\eqn\mwzqcd{M_W = \half g \sqrt{n_G} f_\pi \ts, \qquad
M_Z = \half \sqrt{g^2 + g'^2} \sqrt{n_G} f_\pi \ts,}
while the photon, $A = (g' W^3 + g W^0)/\sqrt{g^2 + g'^2}$, remains
massless. The three Goldstone bosons coupling to the electroweak currents
now appear in the physical spectrum only as the longitudinal components of
the $W^\pm$ and $Z^0$. This is the {\it dynamical} Higgs mechanism.%
\ref\dynhiggs{R.~Jackiw and K.~Johnson, {\it Phys.~Rev.}~{\bf D8}
(1973) 2386\semi
J.~Cornwall and R.~Norton, {\it Phys.~Rev.}~{\bf D8} (1973) 3338\semi
E.~Eichten and F.~Feinberg, {\it Phys.~Rev.}~{\bf D10} (1974) 3254.}
The strong dynamics of QCD did it all; no unnatural, trivial and otherwise
bothersome elementary Higgs bosons were needed.

Unfortunately, this QCD--induced breakdown of electroweak symmetry is
phenomenologically unacceptable. For three generations of quarks and
leptons, Eq.~\mwzqcd\ yields $M_W \cong 52.7\,\mev$ and $M_Z \cong
59.6\,\mev$
\ref\footb{We use electroweak couplings renormalized at 100~GeV. This is
inappropriate for this calculation, but does not introduce significant
numerical error.}.
The measured values of the $W$ and $Z$ masses are 1500 times
larger:\cite{\PDG}
\eqn\mwz{M_W = 80.22 \pm 0.26\,\gev \ts, \qquad M_Z = 91.173 \pm
0.020\,\gev \ts.}
The residual chiral flavor symmetries of quarks and leptons are also wrong.
The electroweak gauge interactions leave invariant a large subgroup of
$\CG_{\chi}$,
\eqn\schi{
\CS_{\chi} = (SU(n_G)_u \otimes SU(n_G)_d \otimes U(1))_L \otimes
           (SU(n_G)_u \otimes SU(n_G)_d \otimes U(1))_R \ts,}
where the subscripts $u,d$ indicate that the symmetry generators act only
on up-- or down--type quarks of the indicated chirality. This symmetry
leaves all up--quarks degenerate and, likewise, all down--quarks. All
charged Goldstone bosons acquire equal masses, while all neutral
Goldstone bosons remain massless. All leptons are strictly massless
in this model as well.

Despite this model's failure to describe the observed breakdown of
electroweak and flavor symmetries, it does produce one phenomenological
success. Up to calculable corrections of $O(\alpha)$,
\nref\marvin{M.~Weinstein, {\it Phys.~Rev.}~{\bf D8} (1973) 2511. To my
knowledge, this is the first paper in which it was recognized
that the condition $\rho = 1$ relies only on the $SU(2)$ quantum
numbers of the symmetry--breaking mechanism, and not on whether
Higgs bosons are elementary scalars.}%
\nref\swtc{S.~Weinberg, {\it Phys.~Rev.}~{\bf D13} (1976) 974;
{\it ibid} {\bf D19} (1979) 1277.}%
\nref\lstc{L.~Susskind, {\it Phys.~Rev.}~{\bf D20} (1979) 2619.}%
\cite{\marvin,\swtc,\lstc}
\eqn\rhoone{\rho \equiv {M_W^2 \over {M_Z^2 \cos^2\theta_W}} = 1 \ts,}
where $\cos\theta_W = g/\sqrt{g^2 + g^{'2}}$. Experimentally, $\rho =
-0.998 \pm 0.003$ (see Ref.~\PDG, ppIII.59, ff and Eq.~(2.8) below).
The basis of this prediction  is easy to understand. The
electroweak--symmetry breaking condensates in \qcond\ leave invariant a
``custodial'' $SU(2) \otimes SU(2)$ subgroup  of $\CG_{\chi}$ which, in
turn, contains $(SU(2) \otimes U(1))_{\rm EW}$.

\medskip
\noindent{\it 1.4 Technicolor}
\smallskip

The solution to the problem of too small a breaking scale for $SU(2)
\otimes U(1)$ is clear:\cite{\swtc,\lstc} Assume that there is a new
asymptotically free gauge interaction, called ``technicolor'', with gauge
group $G_{TC}$, and gauge coupling $\atc$ that becomes strong in the
vicinity of a few hundred~GeV. In simple technicolor models,
one assign $N_D$ doublets of left-- and right--handed technifermions,
$T_{iL,R} = (U_i, D_i)_{L,R}$, to equivalent complex
irreducible representations of $G_{TC}$. If the $T_L$ are assigned to
electroweak $SU(2)$ as doublets and the $T_R$ as singlets, with appropriate
(nonanomalous) $U(1)$ couplings for all the technifermions, then they are
{\it massless} and have the chiral flavor group
\eqn\tgchi{G_\chi = SU(2 N_D)_L \otimes SU(2 N_D)_R \supset SU(2)_L \otimes
SU(2)_R \ts.}

When $\atc$ becomes strong, technifermion condensates form, similar
to those in Eq.~\qcond (technicolor indices are suppressed here):
\eqn\tcond{\langle \Omega \vert \ol U_{i L} U_{j R} \vert \Omega
\rangle = \langle \Omega \vert \ol D_{i L} D_{j R} \vert \Omega
\rangle = - \delta_{ij} \Delta_T \ts .}
\ifx\ctc\undefined\let\ctc\relax\fi
The chiral symmetry breaks to $S_\chi = SU(2 N_D) \supset SU(2)_V$ and
there are $4 N_D^2 - 1$ massless Goldstone bosons with decay constant
$F_{\pi_T}$. Throughout these lecture notes, we assume $\Delta_T \simeq
4 \pi F^3_{\pi_T}$;\cite{\delqcd} see Eq.~\ctc\ below.
A diagonal linear combination of three of these are
absorbed as the longitudinal components of the $W^\pm$ and $Z^0$ weak bosons,
which acquire mass
\eqn\mwztc{M_W = \half g \sqrt{N_D} F_{\pi_T} \ts, \qquad
M_Z = \half \sqrt{g^2 + g'^2} \sqrt{N_D} F_{\pi_T} = M_W/\cos\theta_W \ts.}
Thus, the scale $\Lambda_{TC}$ at which technicolor interactions become
strong and break $G_\chi$ to $S_\chi$ is determined by the weak scale,
$\Few = 246\,\gev$:
\eqn\fpit{\eqalign{
&F_{\pi_T} = \Few/\sqrt{N_D} \ts;\cr
&\Lambda_{TC} = {\rm few} \ts \times \ts F_{\pi_T} \ts. \cr}}

What have we achieved? First, a dynamical explanation for electroweak
symmetry breaking. It is the same phenomenon that causes chiral symmetry
breakdown in QCD. Second, the mechanism is natural and stabilizes the weak
scale far below $M_P$. The technifermion chiral symmetry and
$SU(2)\otimes U(1)$ do not break until $\atc$ becomes large enough that
condensates $\langle \ol T_L T_R \rangle$ form. Since technicolor is an
asymptotically free interaction, it is natural to suppose that $\atc$ is
small at very high scales ($O(10^{16}\,\gev)$, say), and then grows to
become strong as we descend in energy to $O(1\,\tev)$. Since the chiral
symmetry breaking is a soft phenomenon, vanishing rapidly at energies above
$\Lambda_{TC}$ in an asymptotically free theory,%
\ref\kdlhdp{K.~Lane, {\it Phys.~Rev.}~{\bf D10} (1974) 2605\semi
H.~D.~Politzer {\it Nucl.~Phys.}~{\bf B117} (1976) 397.}
all mass scales associated with the breaking are of $O(\Lambda_{TC})$.
Third, the theory is nontrivial. Because the technicolor $\beta$--function
is always negative, $\atc$ does not develop a Landau pole as did the Higgs
scalar self--coupling in Eq.~\lamtriv. Fourth, as in the elementary Higgs
model, there is, quite naturally, a custodial $SU(2)_R$ flavor symmetry in
$G_\chi$ and this guarantees $\rho = M_W^2 /M_Z^2 \cos^2\theta_W = 1 +
O(\alpha)$. This custodial symmetry was a consequence of the assignment of
$T_L$ and $T_R$ to equivalent, complex representations of $G_{TC}$.

What we have not achieved is an explanation of quark and lepton flavor
symmetries, much less a theory of flavor symmetry breaking. The quarks and
leptons of the technicolor theory constructed here remain massless. We
shall return to this problem soon.

\vfil\eject
%\medskip
\noindent{\it 1.5 Minimal Technicolor}
\smallskip

The minimal technicolor model has one doublet of technifermions, $T_{L,R}$.
Assigning them to equivalent complex representations of $G_{TC}$, their
chiral symmetry $G_\chi = SU(2)_L \otimes SU(2)_R$ breaks
down to $S_\chi = SU(2)_V$. If $T_L$ and $U_R$, $D_R$ are assigned to
electroweak $SU(2) \otimes U(1)$ as $({\bf 2},0)$ and $({\bf 1},\half)$,
$({\bf 1},-\half)$, respectively, all gauge interactions are anomaly--free.
The three Goldstone bosons $\pi_T^\pm$ and $\pi_T^0$ are absorbed into
$W_L^\pm$ and $Z_L^0$.%
\ref\footc{Strictly speaking the combinations absorbed in the Higgs
mechanism are the Goldstone bosons coupling to the weak axial currents.
These are $(\Few \pi_T + f_\pi \pi_q)/\sqrt{\Few^2 + f_\pi^2}$, where
$\pi_q$ is an appropriate linear combination of Goldstone bosons in the
quark sector.}

Just as happens in QCD, this model should
have an infinite tower of bound states---technihadrons---that can
be classified according to $SU(2)_V$. Most
notably, there will be an isotriplet of vector mesons, $\tro$, and an
isosinglet $\omega_T$. To estimate their masses, it is customary to suppose
that $G_{TC} = SU(N_{TC})$ and $T_{L,R}$ belongs to the fundamental
representation~${\bf N_{TC}}$. Then, it is assumed that one can use
large--$N_{TC}$
arguments to scale masses and decay couplings from the $\rho$ and
$\omega$ of QCD. This gives the mass\cite{\ehlq}
\eqn\trhomass{M_{\tro} \cong M_{\omega_T} =
\sqrt{{3 \over {N_{TC}}}} \ts {F_{\pi_T} \over {f_\pi}} \ts M_\rho \simeq
2\sqrt{{3 \over {N_{TC}}}}\,\tev \ts.}
As in QCD, $\tro$ and $\omega_T$ decay into technipions, which are the
$W_L^\pm$ and $Z_L^0$. Again using a large--$N_{TC}$ argument
and scaling from QCD,
\eqn\trhowidth{\eqalign{
& \Gamma(\tro \ra W_L W_L) = {3 \over {N_{TC}}} \ts {M_{\tro} \over {M_\rho}}
\ts {\Gamma(\rho \ra \pi\pi) \over {v_\pi^3}}
\simeq 500 \left({3\over{N_{TC}}} \right)^{3/2}\,\gev  \ts; \cr
& \Gamma(\omega_T \ra W_L^+ W_L^- Z_L^0) = \left({3\over{N_{TC}}} \right)^2
\ts {M_{\omega_T} \over {M_\omega}} \ts
{\Gamma(\omega \ra \pi^+\pi^- \pi^0) \over
{{{\rm phase \ts\ts space}}}} \simeq 80 \left({3\over{N_{TC}}}
\right)^{5/2}\,\gev \ts. \cr}}
Here $v_\pi = \sqrt{1 - 4 M_\pi^2/M_\rho^2}$.
Because $\tro$ and $\omega_T$ are so much heavier than the $W$ and $Z$,
it is possible that they have appreciable decay rates
into states with more than
two or three weak bosons. To my knowledge, no one has attempted
an estimate of these rates.

Because they are color singlets, the $\tro$ and $\omega_T$ are produced
only weakly in hadron and $e^+ e^-$ collisions via weak--boson dominated
production and weak boson fusion.%
\nref\snowmass{ K.~Lane, Proceedings of the 1982 DPF Summer Study
on  Elementary Particle Physics and Future Facilities, edited by
R.~Donaldson, R.~Gustafson and F.~Paige (Fermilab 1983), p.~222.}%
\nref\bulos{F.Bulos, et al., Proceedings of the 1982 DPF Summer Study
on  Elementary Particle Physics and Future Facilities, edited by
R.~Donaldson, R.~Gustafson and F.~Paige (Fermilab 1983), p.~71.}
\nref\rscWW{R.~S.~Chivukula, {\it Proceedings of the $12^{th}$ Johns
Hopkins Workshop on Current  Problems in Particle Theory} (Baltimore, MD,
June 8--10, 1988), G.~Domokos and S.~Kovesi--Domokos, eds., World Scientific
(Singapore, 1988).}%
\cite{\ehlq,\snowmass,\rscWW} If the mass and width estimates above are
correct and $N_{TC}$ is not large, the only hope for detecting the $\tro$
is at an SSC--class hadron collider or at a 2~TeV $e^+e^-$ collider. In
either case, an integrated luminosity of about $\uhly = 100\ts \fb^{-1}$
would be required for discovery. It is unlikely that the $\omega_T$ could
be reconstructed in its $W_L^+ W_L^- Z_L^0$ decay decay channel. Here, the
only hope is to use the rare decay $\omega_T \ra Z_L^0 \gamma$, estimated
to have a rate of only a few per~cent of the $WWZ$ mode.%
\ref\rscomegaT{R.~S.~Chivukula and M.~Golden, {\it Phys.~Rev.}~{\bf D41},
(1990) 2795.}
Other technimesons of interest are the $a_{1T}$, an isovector axial--vector
meson analogous to the $a_1(1260)$ and a isosinglet scalar
technimeson, $f_{0T}$, analogous to the broad $f_0(1400)$. This latter
object, with mass between one and two~TeV, is the nearest approximation to
the standard Higgs boson, $\h$. Neither of these mesons have received much
phenomenological study. One would also expect technibaryons. These would be
fermions or bosons, depending on the $G_{TC}$--representation content of the
technifermions.

\medskip
\noindent{\it 1.6 The One--Family Technicolor Model}
\smallskip

The next simplest technicolor model has one complete family of
technifermions---a doublet of color--triplet ``techniquarks'' and a doublet
of color--singlet ``technileptons''%
\nref\farhi{E.~Farhi and L.~Susskind {\it Phys.~Rev.}~{\bf D20}
(1979) 3404.}%
\cite{\ehlq,\snowmass,\farhi}
\eqn\tdoub{\eqalign{
&Q_{L,R} = \pmatrix{U\cr D\cr}_{L,R} \in {\bf 3} \ts\ts\ts {\rm of}
\ts \ts\ts SU(3) \cr\cr
&L_{L,R} = \pmatrix{N\cr E\cr}_{L,R}  \in {\bf 1} \ts\ts\ts {\rm of}
\ts \ts\ts SU(3) \ts. \cr}}
The chiral techniquarks and technileptons are assumed to transform
according to the same complex representation of $G_{TC}$. Then, all
interactions are anomaly--free the $Q_{L,R}$ and $L_{L,R}$ are assigned
the same $SU(2)\otimes U(1)$ quantum numbers as quarks and leptons.

Since the QCD coupling is relatively weak at the technicolor scale,
$\Lambda_{TC}$, the approximate chiral symmetry of this model is $G_\chi =
SU(8)_L \otimes SU(8)_R$.  When the $SU(3)$--preserving condensates
$\langle U_L U_R \rangle = \langle D_L D_R \rangle \cong  \langle N_L N_R
\rangle = \langle E_L E_R \rangle = -\Delta_T$ form, this symmetry breaks
down to $SU(8)_V$ with 63~Goldstone bosons. These may be classified
according to their transformation properties under custodial $SU(2)_V$ and
color--$SU(3)$ as follows:
\eqn\tpions{\eqalign{
&\tpi^{\pm,0} = W_L^{\pm, 0} \in (\bf 3, \bf 1) \cr
&P^{\pm,0} \in (\bf 3, \bf 1) \cr
&P^{'0} \in (\bf 1, \bf 1) \cr
&\tpiqq \in (\bf 3 , \bf 8) \oplus (\bf 1, \bf 8) \cr
&\tpilq \in (\bf 3, \bf 3) \oplus (\bf 1, \bf 3)\ts, \qquad
\tpiql \in (\bf 3, \bf 3^*) \oplus (\bf 1, \bf 3^*) \ts. \cr}}
The decay constant of these technipions is $F_{\pi_T} =
246\,\gev/\sqrt{N_D} = 123\,\gev$.

There will also be a set of 63~$\tro$ (as well as one $\omega_T$) having
the same $SU(2)_V \otimes SU(3)$ quantum numbers and decaying into pairs of
these technipions. Assuming again that $Q_{L,R}$ and $L_{L,R} \in {\bf N_{TC}}$
of $G_{TC} = SU(N_{TC})$ and that scaling from QCD is permissible,
\eqn\thadrons{\eqalign{
&M_{\rho_T} = \sqrt{{ 3 \over {N_{TC}}}} {F_{\pi_T} \over {f_\pi}} M_\rho
\simeq
\sqrt{{ 3 \over {N_{TC}}}}\,\tev \cr
\sum \Gamma(\tro \ra \tpi\tpi) &\simeq
4 \ts {3 \over {N_{TC}}} \ts {M_{\rho_T} \over {M_\rho}}
\ts {\Gamma(\rho \ra \pi\pi) \over {v_\pi^3}}
%\cr & \qquad\qquad
\simeq 1000 \left({3\over{N_{TC}}}\right)^{3/2}\,\gev \ts. \cr}}
The color--singlet $\tro$ with the same quantum numbers as those in the
one--doublet model are produced weakly in hadron and $e^+e^-$ colliders, as
described above. While their masses are half that expected in the minimal
model, their widths are much greater because of the many open decay
channels (see below). The color--singlet $\tro$ signals would therefore be
broad, difficult--to--see enhancements in $\pi_T$ pair--production. The chances
for discovery of the electrically neutral color--octet $\tro$ are more
promising because they are copiously produced in hadron colliders via their
coupling to a single gluon.\cite{\ehlq}

In Eq.~\thadrons, it is assumed that $M_{\pi_T} \ll M_{\rho_T}$. This is
certainly true in the one--family model described here. The only sources of
explicit $G_\chi$ breaking are the $SU(3) \otimes SU(2) \otimes U(1)$
interactions. These generate the following technipion masses, calculated by
standard current--algebraic techniques and by scaling from QCD where
necessary:%
\nref\eekletc{E.~Eichten and K.~Lane, {\it Phys.~Lett.}~{\bf 90B}
(1980) 125.}%
\nref\vacalign{V.~Baluni, {\it Ann.~Phys.}~{\bf 165} (1985) 148\semi
M~E.~Peskin, {\it Nucl.~Phys.}~{\bf B175} (1980) 197\semi
J.~P.~Preskill, {\it Nucl.~Phys.}~{\bf B177} (1981) 21.}%
\nref\sdim{S.~Dimopoulos, {\it Nucl.~Phys.}~{\bf B168} (1980) 69.}%
\cite{\eekletc,\vacalign,\sdim}
\eqn\tpimass{\eqalign{
&M(P^0)= M(P^{'0}) \simeq 0 \cr
&M(P^\pm) \simeq 7\,\gev \cr
&M(\tpiqq) \simeq 275 \sqrt{3/N_{TC}}\,\gev \cr
&M(\tpilq) \simeq 185 \sqrt{3/N_{TC}}\,\gev \ts. \cr}}
The $P^\pm$ mass arises from electroweak interactions, while the
color--octet and triplet technipions' masses are due to color--$SU(3)$.

The technicolor models described so far are unacceptable. Quarks and leptons
are still massless because nothing explicitly breaks their chiral
symmetries. The charged technipions $P^\pm$ would have been discovered long
ago in $e^+e^-$ annihilation and $Z^0$ decay if they existed with such
small masses.\cite{\eekletc} The nearly massless $P^0$ and $P^{'0}$ are
like the Weinberg--Wilczek axion.%
\ref\axion{S.~Weinberg, {\it Phys.~Rev.~Lett.}~{\bf 40} (1978) 223\semi
F.~Wilczek, {\it Phys.~Rev.~Lett.}~{\bf 40} (1978) 279.}
They couple to ordinary matter with a strength of $O(M_q/F_{\pi_T})$, where
$M_q$ is a QCD--generated dynamical mass, and they decay to two photons. They,
too, are likely to have been ruled out by the standard axion
searches.\cite{\PDG} Curing these problems is the motivation for
extended technicolor.

\vfil\eject
%\medskip
\noindent{\it 1.7 Extended Technicolor---A Dynamical Scenario for Flavor
Physics}
\smallskip

We have just seen that a theory with only technicolor and color strong
interactions leaves too much chiral symmetry. Most of these symmetries are
spontaneously broken when the gauge couplings become large, but they are
not {\it explicitly} broken. As a consequence, quarks and leptons have no
{\it hard} masses. Quarks get a dynamical mass $M_q = O(300\,\mev)$ from
QCD, but that's it. The pions, kaons, and eta are massless or nearly so.
Leptons are strictly massless. (Imagine what atomic and nuclear physics are
like!) Technifermions do not acquire hard masses either. Those
technifermion chiral symmetries that commute with the standard--model gauge
interactions are unbroken, so there are axion--like $P^0$ and $P^{'0}$ and
light $P^\pm$.

Once the problem is stated in this form, the solution is obvious: introduce
new interactions that break the unwanted symmetries. Assume that these
new interactions occur at energies well below the Planck scale. Then, in the
spirit of dynamical electroweak symmetry breaking, they should be gauge
interactions, involving fermions as the only matter fields. To break the
quark, lepton and technifermion flavor symmetries, we must gauge all or
part of these symmetries. This means putting quarks, leptons and
technifermions together into the same (irreducible) representations of the
new gauge group. This new gauge group must then contain both technicolor
and flavor. The interaction is now called ``extended technicolor'' (ETC,
for short).%
\nref\sdlsetc{S.~Dimopoulous and L.~Susskind, {\it Nucl.~Phys.~}~{\bf B155}
(1979) 237.}%
\cite{\eekletc,\sdlsetc}

The ETC gauge interactions involve currents coupling $T$, $q$, $\ell$ in
such a way that the only symmetries still intact just above the electroweak
scale are $G_{TC} \otimes SU(3) \otimes SU(2) \otimes U(1) \otimes B
\otimes L$. As we shall discuss in more detail below, the only way we know
to do this with any degree of economy is to embed technicolor, color and
part of electromagnetic $U(1)$ into the gauge group $\Getc$.\cite{\eekletc}
This severely constrains ETC model--building.

At the high scale $\Lambda_{ETC} \gg \Lambda_{TC}$, the ETC gauge symmetry
breaks down to $G_{TC} \otimes SU(3) \otimes \cdots$. Exactly how this is
done is left unspecified in almost all ETC models. After the breaking,
there are heavy ETC gauge bosons, with mass $\Metc \sim g_{ETC}
\Lambda_{ETC}$, where $g_{ETC}$ is a generic ETC gauge coupling
renormalized at $\Lambda_{ETC}$. We shall suppose that $g_{ETC}$ is not
much less than one.

Quark and lepton hard masses are generated in $O(g^2_{ETC})$ by a light
fermion's turning into a technifermion and back into a (possibly different)
light fermion while emitting and reabsorbing a heavy ETC gauge boson. The
required change of light fermion chirality is induced by the
technifermion's dynamically--generated mass. The $q$, $\ell$ self--energy
graphs may be estimated by using the operator product expansion. The typical
momentum running around the self--energy loop is $\Metc$. The generic
result is, ignoring mixing angles and the fact that different ETC couplings
and boson masses may contribute to $m_q$ and $m_\ell$,
\eqn\qmass{m_q(\Metc) \simeq m_\ell(\Metc)  \simeq {g_{ETC}^2 \over
{M_{ETC}^2}} \condetc \ts.}
Here, we have noted the fact that these masses are renormalized at $\Metc$,
as is the condensate $\condetc = \langle \Omega \vert \ol T T \vert \Omega
\rangle_{M_{ETC}}$. This condensate is related to the one renormalized at
$\Lambda_{TC}$, expected by scaling from QCD to be
\eqn\ctc{\condtc \simeq 4 \pi F^3_{\pi_T} \ts,}
by
\eqn\keyeq{
\condetc = \condtc \ts \exp\left(\int_{\Lambda_{TC}}^{M_{ETC}} \ts {d \mu
\over {\mu}} \ts \gmm(\mu) \right) \ts.}
Here,
\eqn\gamm{\gmm(\mu) = {3 C_2(R) \over {2 \pi}} \atc(\mu) + O(\atc^2) \ts,}
is the anomalous dimension of the operator $\ol T T$ and $C_2(R)$ is
the quadratic Casimir of the technifermion $G_{TC}$--representation $R$.
Technifermion $G_{TC}$ and flavor indices have been suppressed in
Eqs.~\keyeq\ and \gamm, but they are easily included if necessary. If
technifermions acquire a mass from ETC interactions, $m_T(\Metc)$ is
given by a similar equation. This ``hard'' mass runs according to the
equation
\eqn\runmass{
m_T(\Metc) = m_T(\Lambda_{TC}) \ts \exp\left(-\int_{\Lambda_{TC}}^{M_{ETC}}
\ts {d \mu \over {\mu}} \ts \gmm(\mu) \right) \ts.}
Thus, $m_T \ts \ol T T$ is a renormalization--group invariant.
Equations~\qmass, \keyeq\ and \gamm\ are the key equations of extended
technicolor. Together with Eq.~\ctc, they may be used to
estimate most quantities of phenomenological interest, including
$\Lambda_{ETC}$ and the ETC--generated masses of technipions.

Let us estimate $\Lambda_{ETC}$ for $m_q = 1\,\gev$. Since color--$SU(3)$ is
a relatively weak interaction above 100~GeV, we can ignore the running of
the quark mass according to its $\gmm$ and set $m_q(1\,\gev) \cong
m_q(\Metc)$. If technicolor is like QCD, that is, it is precociously
asymptotically free above $\Lambda_{TC}$ so that $\atc(\mu)$ is also small
in the integral in Eq.~\keyeq, then we can also ignore the running of the
technifermion condensate. Then,
\eqn\lametc{
\Lambda_{ETC} \equiv {M_{ETC} \over {g_{ETC}}} \simeq \sqrt{{4 \pi F_\pi^3
\over {m_q N_D^{3/2}}}} \simeq 14 \sqrt{{1\,\gev \over{m_q\ts N_D^{3/2}}}} \ts
\,\tev \ts.}

The generic ETC contribution to technipion masses is calculated in the
usual current--algebraic (or chiral Lagrangian) way, using an
effective interaction of the form
\eqn\efflag{
{g^2_{ETC} \over {\Metc^2}} \ts \ol T_L
\gamma^\mu T_L \ts \ol T_R \gamma_\mu T_R = m_T \ts \ol T T }
for the explicit chiral--symmetry breaking.%
\ref\footd{It is really not proper to integrate out the technifermions at
the scale $\Lambda_{TC}$ as is done in Eq.~\efflag. Nevertheless, it is
convenient to use the interaction $m_T \ol T T$ to calculate the technipion
masses and one gets the correct result by this method.}
The result is
\eqn\tpimasssq{
F^2_{\pi_T} M^2_{\pi_T}  \simeq 2\ts {g^2_{ETC} \over {M^2_{ETC}}}
\ts \langle \ol T_L T_R \ol T_R T_L \rangle_{ETC} \simeq 2\ts m_T(M) \ts
\langle \ol T T \rangle_M \ts.}
For $\gmm$ small and $m_T(\Lambda_{TC}) \cong m_T(\Metc)$, we get
\eqn\tpimass{
M_{\pi_T}  \ts\simeq 55 \ts \sqrt{{m_T(\Lambda_{TC})
\over {1\,\gev \times N_D}}}\ts \gev \ts.}
This mass, added in quadrature with contributions from other sources,
may be large enough to remove the phenomenological objections to
technipions.

Extended technicolor is a dynamical theory of flavor at moderate energies,
at least compared to $M_P$. One can easily imagine that a complete
ETC theory would have {\it no} free parameters other than the value of some
gauge coupling at very high energies. Just set the theory off at this high
energy and, as we descend, the gauge dynamics would break all the
symmetries of the theory other than $G_{TC} \otimes SU(3) \otimes
U_{EM}(1)$. Such a theory almost certainly would have effects (technipions,
e.g.) discernible at existing or still--planned colliders.

We are far from this ideal. We do not even have a compelling model,
though much effort has been put into trying to build them.%
\ref\etcmodels{To name a few examples: E.~Farhi and L.~Susskind in
Ref.~\farhi\semi
S.~Raby, S.~Dimopoulos and L.~Susskind, {\it Nucl.~Phys.}~{\bf B169}
(1980) 373\semi
S.~Dimopoulos, H.~Georgi and S.~Raby, {\it Phys.~Lett.}~{\bf 127B}
(1983) 101\semi
S.--C.~Chao and K.~Lane, {\it Phys.~Lett.}~{\bf 159B} (1985) 135\semi
S.~King and S.~Mannan, {\it Nucl.~Phys.}~{\bf B369} (1992) 119\semi
M.~Einhorn and D.~Nash, {\it Nucl.~Phys.}~{\bf B371} (1992) 32\semi
R.Sundrum, {\it Nucl.~Phys.}~{\bf B395} (1993) 60\semi
L.~Randall, {\it Nucl.~Phys.}~{\bf B403} (1993) 122\semi
T.~Appelquist and J.~Terning, Yale and Boston University Preprint,
YCTP--P21--93, BUHEP--93--23 (1993).}
Even quite general questions have no definite answer yet:

\item{$\bullet$} What breaks ETC?
Since the only relevant interactions at $\Lambda_{ETC}$ are gauge
interactions, ETC breaking presumably is a dynamical Higgs process. What
interactions are responsible for {\it that} Higgs mechanism?
\ref\foote{Atlas holds up the world. He stands on the back of a great
turtle. What does the turtle stand on? {\it It's turtles all the way down!}}

\item{$\bullet$} Exactly what gives rise to the large hierarchy of
quark and lepton masses
that we observe? Do the ETC gauge interactions ``tumble'', breaking down at
a succession of scales, with the lightest generation peeling off at the
highest scale and getting mass from suppressed ETC interactions at this
scale, and so on?%
\ref\tumble{S.~Raby, S.~Dimopoulos and L.~Susskind, {\it Nucl.~Phys.~}~{\bf
B169} (1980) 373\semi V.~Baluni, unpublished.}
Or, are the quark and lepton hierarchies a consequence of diagonalizing
some large ETC boson mass matrix?%
\ref\footf{If the anomalous dimensions are
small, $m_{q, \ell}$ scale as $1/\Metc^2$. Then, an ETC hierarchy of about
two orders of magnitude is required to span the range from $m_u$ to
$m_t$.}

\item{$\bullet$} The top--quark mass is known to be greater than about
100~GeV.\cite{\topmass} According to Eq.~\lametc, this large mass is
generated by an ETC scale of about 1~TeV. This is essentially the same as
the technicolor scale---a nonsensical result. So, how do we explain the
top mass in ETC?

\medskip
\noindent{\it 1.8 An Important Constraint on Extended Technicolor}
\smallskip

Let us assume that the technicolor group $G_{TC}$ is simple. As far as I
know, model--builders always assume this---usually implicitly. It means,
among other things, that the technifermions from which quarks and leptons
get their hard masses all have the same technicolor interactions. In
particular, if techniquarks and technileptons exist, as in the one--family
model, they transform according to the same technicolor group. This
assumption is crucial to the argument below. Let us also assume that the
ETC gauge interactions are electroweak--$SU(2)$ invariant, i.e., $[G_{ETC},
SU(2)_{EW}] = 0$. This assumption is also commonly made---for simplicity.
I do not know if it is essential to the argument below; I suspect that it
is not.

It follows from these assumptions and the absence of ``classical''
Weinberg--Wilczek axions that {\it all} known fermions---quarks and leptons
and the technifermions to which they couple---must belong to {\it {\ub {at
most three irreducible representations of}}} {\it {\ub
{$G_{ETC}$}}}.\cite{\eekletc} These representations are:

\medskip

\item{\bull} The equivalent representations $\CU_L \sim \CD_L$, containing
$q_{iL} = (u_{iL}, d_{iL})$, $\ell_{iL} = (\nu_{iL}, e_{iL})$ and $T_L =
(T_{U_L}, T_{D_L})$ ($T--$flavor labels are suppressed).

\medskip

\item{\bull} $\CU_R$ containing $u_{iR}$, $\nu_{iR}$ (if they exist) and
$T_{U_R}$.

\medskip

\item{\bull} $\CD_R$ contains $d_{iR}$, $e_{iR}$ and $T_{D_R}$.

\medskip

\noindent The equivalence of $\CU_L$ and $\CD_L$ is required by our
assumption that $[G_{ETC}, SU(2)_{EW}] = 0$. One of $\CU_R$ and $\CD_R$ may
be equivalent to $\CU_L$. But, in order that the up-- and down--quark mass
matrices are not equal and the Kobayashi--Maskawa matrix is not trivial,
$\CU_R$ and $\CD_R$ must be {\it inequivalent}. If there are more ETC
irreducible representations containing the known fermions than these, it is
always possible to construct a $U(1)$ symmetry current involving
technifermions, quarks, and leptons and which is conserved up to a
color--$SU(3)$ anomaly.%
\ref\footg{This is where the assumption of a simple $G_{TC}$ comes in.
If $G_{TC}$ is not simple, it may not be possible to cancel the strong
technicolor anomalies in this current's divergence.}
This symmetry is spontaneously broken when the technifermions condense,
generating a very light pseudo--Goldstone boson. The boson's
mass is of order $\Lambda_{QCD}^2/\Lambda_{TC} \sim 100~{\rm KeV}$ and its
couplings to ordinary fermions of order $m_f/F_{\pi_T}$;
this is a ``classical'' axion.

As a first corollary, $SU(3)_C$ must be embedded in $G_{ETC}$. If it were
not, then quarks and leptons would get their mass from different sets of
technifermions, color--triplets $Q$ and color--singlets $L$, respectively.
The ``axion'' $U(1)$ current would then be given by $\ol Q \gamma_\mu
\gamma_5 Q - 3\ts [T(Q)/T(L)] \ts \ol L \gamma_\mu \gamma_5 L$, where
$T(Q)$ and $T(L)$ are the trace of the square  $G_{TC}$--representation for
$Q$ and $L$. Thus, as stated earlier, at the scale(s) $\Lambda_{ETC}$, the
symmetry $G_{ETC} \ra G_{TC} \otimes SU(3)_C \otimes \cdots$ and, so, there
is at least a partial unification of gauge interactions above this scale.

Since quarks and leptons must coexist in the same ETC representations,
it follows that the ETC gauge group cannot commute with electric
charge. Then, since we assumed $\left[G_{ETC},\ts SU(2)_{EW} \right] = 0$,
we have as a second corollary that at least a piece of the usual
electroweak hypercharge group, $U(1)_{EW}$, must be embedded in $G_{ETC}$.

Other important constraints imposed by mathematical consistency and
phenomenology are:%
\ref\KLunpub{K.~Lane, seminar given at Yale University, 1986; unpublished.}

\medskip

\item{1.} The ETC interactions should be asymptotically free. Otherwise,
naturalness is lost and, in particular, it is difficult to understand
spontaneous symmetry breaking.

\item{2.} There must be no gauge anomalies.

\item{3.} Flavor--changing neutral currents must not be unacceptably large.
We'll have more to say about this soon.

\item{4.} The top--quark mass must be greater than 100~GeV (unless it decays
as $t \ra b \pi^+_T$ with $M_{\pi_T} \simge 45\,\gev$).\cite{\topmass}

\item{5.} The parameter $\rho = 1 + O(\alpha)$. This may be difficult to
implement in an ETC model which accomodates the large top--bottom splitting.

\item{6.} The masses, if any, of the usual neutrinos must be acceptably
small.

\item{7.} There must be weak $CP$--violation without strong $CP$--violation
(i.e., $\ol \theta = \theta_{QCD} + \arg \det m_q \simle 10^{-8}$) and
without visible axions.

\item{8.} There must be no extra ``photons''---massless (or very light)
gauge bosons.

\item{9.} Quarks and leptons must have the proper electric charges.

\medskip

\noindent These constraints and the one on $G_{ETC}$ representations make
ETC model building a very difficult enterprise. Indeed, no really
satisfactory ETC model has yet been constructed. Most model--builders are
content to invent toy models that illustrate a particular new idea or
trick. This is a reflection of how hard the flavor problem is. Clearly, we
need experimental input.

The difficulty of building realistic ETC models that are relatively simple
and compelling is, I believe, the real reason for technicolor's
unpopularity. Other, more popular, approaches (such as elementary Higgs
boson models, with or without supersymmetry) appear simple and attractive
because they do not attempt to explain the physics of flavor---it is
postponed to the highest energy scales. Flavor physics is hard!

\newsec{WHY NOT TECHNICOLOR?}

In this section we review the principal phenomenological objections to
technicolor---the reasons why so many people say that ``technicolor is
dead''. As I just said, however, I think the real reason people say this is
that model--building is so difficult.

\medskip
\noindent{\it 2.1 Flavor--Changing Neutral Currents}
\smallskip

Extended technicolor interactions are expected to have flavor--changing
neutral currents (FCNC) involving quarks and, probably, leptons. The reason
for this is simple: Realistic quark mass matrices require ETC transitions
between different flavors: $q \ra T \ra q'$. But the algebra of ETC
generators of the form $\ol q' \gamma_\mu T$ and $\ol T \gamma_\mu q$
must include the
generator $\ol q' \gamma_\mu q$. After diagonalization of the
quark mass matrices, this will produce a flavor--changing neutral current
coupling to an ETC boson whose mass is not likely to be much different from
the masses of those coupling to $\ol q' \gamma_\mu T$ and $\ol T
\gamma_\mu q$. Thus, in general, we expect four--quark interactions,
mediated by ETC boson exchange, and these can generate highly forbidden
process. The same argument leads one to expect two--quark--plus--lepton and
four--lepton interactions that may generate unobserved processes.%
\nref\ellisfcnc{J.~Ellis, M.~Gaillard, D.~Nanopoulos and P.~Sikivie,
{\it Nucl.~Phys.~}~{\bf B182} (1981) 529.}%
\cite{\eekletc,\ellisfcnc} Despite some effort, no satisfactory
GIM--like mechanism%
\ref\GIM{S.~L.~Glashow, J.~Iliopoulos and L.~Maiani,
{\it Phys.~Rev.}~{\bf D2} (1970) 1285\semi
S.~L.~Glashow and S.~Weinberg, Natural Flavor Conservation,
{\it Phys.~Rev.}~{\bf D15} (1977) 1958.}
has been found to eliminate these FCNC interactions.%
\ref\ETCGIM{See the papers by S.~Dimopoulos, H.~M.~Georgi and S.~Raby;
S.--C.~Chao and K.~Lane; and L.~Randall in Ref.~\etcmodels.}

The most stringent constraint on ETC comes from $\vert \Delta S \vert = 2$
interactions. Such an interaction has the generic form
\eqn\dstwo{\CL_{\vert \Delta S \vert = 2} = {g^2_{ETC} \ts \theta^2_{sd} \over
{M^2_{ETC}}} \ts\ts \ol s \Gamma^\mu d \ts\ts \ol s \Gamma'_\mu d + {\rm
h.c.}}
Here, $\theta_{sd}$ is a mixing--angle factor, presumed to enter twice in
writing the currents in terms of mass eigenstates. It may be complex and it
seems unlikely that it would be much smaller in magnitude than the Cabibbo
angle, say $0.1 \simle \vert \theta_{sd} \vert \simle 1$. The matrices
$\Gamma_\mu$ and $\Gamma'_{\mu}$ are left-- and/or right--chirality Dirac
matrices. The contribution of this interaction to the $K_L - K_S$ mass
difference can be estimated as follows:
\eqn\klks{\eqalign{
2 \ts M_K^0 \ts (\Delta M_K)_{ETC} &= {g^2_{ETC} \ts \theta^2_{sd} \over
{M^2_{ETC}}} \ts \langle \ol K^0 \vert \ol s \Gamma^\mu d \ts \ol s
\Gamma'_\mu d \vert K^0 \rangle + {\rm c.c} \cr
&\simeq {g^2_{ETC} \ts {\rm Re}(\theta^2_{sd}) \over {2 M^2_{ETC}}} \ts
f^2_K M^2_K \ts, \cr}}
where I put
$\Gamma_\mu, \ts \Gamma'_\mu = \half \gamma_\mu \ts (1 - \gamma_5)$
and used the vacuum insertion approximation with
$\langle \Omega \vert \ol s \gamma_\mu \gamma_5 \vert K^0(p) \rangle
= i f_K p_\mu$ and $f_K \simeq 100\,\mev$. This ETC contribution must
be less than the measured mass difference,
$\Delta M_K = 3.5 \times 10^{-12}\,\mev$. This gives the limit
\eqn\etclimit{
{M_{ETC} \over {g_{ETC} \ts \sqrt{{\rm Re}(\theta^2_{sd})}}} > 600\,\tev
\ts.}
If $\theta_{sd}$ is complex, $\CL_{\vert \Delta S \vert = 2}$ contributes
to the imaginary part of the $K^0 - \ol K^0$ mass matrix and the limit is
at least an order of magnitude more stringent.

Assume that $\theta_{sd}$ is real and positive. The fermion hard masses
that come arise from the large scale in \etclimit\ are
\eqn\fcnchard{
m_{q, \ell, T}(M_{ETC}) \simeq {g_{ETC}^2 \over {M_{ETC}^2}}
\condetc < {0.5\,\mev\over{N_D^{3/2} \ts \theta_{sd}^2}} \ts,}
where I used Eq.~\ctc\ to estimate the condensate. The ETC contribution to
the mass of the color--singlet technipions occurring in the one--family and
similar models is
\eqn\etcpimass{
M(P^\pm, P^0, P^{'0}) \simeq \sqrt{{m_T \ts \langle \ol T T \rangle \over
{F^2_{\pi_T}}}} < {1.3\,\gev \over{(\theta_{sd} \ts N_D)}} \ts.}
These are the mass estimates that lead to the familiar statement
``technicolor is dead''.

\medskip
\noindent{\it 2.2 Precision Electroweak Tests}
\smallskip

The standard $SU(2) \otimes U(1)$ model of electroweak interactions has
passed all experimental tests it has faced so far.\cite{\smtests} The
parameters of this model---$\alpha(M_Z)$, $M_Z$, $\sin^2 \theta_W$---are
so precisely known that they may be used to limit new physics at energy
scales above 100~GeV.%
\ref\pettests{B.~W.~Lynn, M.~E.~Peskin and R.~G.~Stuart, in Trieste
Electroweak 1985, (1985) 213;
M.~E.~Peskin and T.~Takeuchi, Phys. Rev. Lett. {\bf 65},
(1990) 964\semi A.~Longhitano, {\it Phys.~Rev.}~{\bf D22} (1980) 1166; {\it
Nucl.~Phys.}~{\bf B188} (1981) 118\semi
R.~Renken and M.~Peskin, {\it Nucl.~Phys.}~{\bf B211} (1983) 93\semi
M.~Golden and L.~Randall, {\it Nucl.~Phys.}~{\bf B361} (1990) 3\semi
B.~Holden and J.~Terning, {\it Phys.~Lett.}~{\bf 247B} (1990) 88\semi
A.~Dobado, D.~Espriu and M~J.~Herrero, {\it Phys.~Lett.}~{\bf 255B} (1990)
405\semi H.~Georgi, {\it Nucl.~Phys.}~{\bf B363} (1991) 301.}
The quantities most sensitive to new physics are defined in terms of
of the correlation functions of electroweak currents as follows: The
correlators have the form
\eqn\pifcn{\int d^4x \ts e^{-i q\cdot x} \langle\Omega | T\left(j^\mu_i(x)
j^\nu_j(0)\right) | \Omega \rangle =
i g^{\mu\nu} \Pi_{ij}(q^2) + q^\mu q^\nu \ts {\rm terms} \ts.}
New, high--mass physics affects the $\Pi_{ij}$ functions. Assuming that the
scale of such new physics is well above $M_{W,Z}$, the so--called {\it
oblique} correction factors $S$, $T$ and $U$ that measure this
new physics are given by
\eqn\stu{\eqalign{
&S= 16\pi {d \over {d q^2}} \left[ \Pi_{33} (q^2) - \Pi_{3Q}(q^2)
\right]_{q^2=0} \ts \equiv \ts 16\pi \left[ \Pi_{33}^{'}(0) - \Pi_{3Q}^{'}(0)
\right] \ts , \cr
&T= {4\pi \over{M^2_Z \cos^2\theta_W \sin^2\theta_W}}
\ts \left[ \Pi_{11}(0) - \Pi_{33}(0) \right] \ts , \cr
&U= 16\pi \left[ \Pi_{11}^{'}(0) - \Pi_{33}^{'}(0)
\right] \ts . \cr}}
The parameter $S$ is a measure of the splitting between $M_W$ and $M_Z$
induced by weak--isospin conserving effects; the $\rho$--parameter
is given by $\rho = 1 + \alpha T$; The $U$--parameter measures weak--isospin
breaking in the $W$ and $Z$ mass splitting. The experimental limits on
$S,T,U$ are a matter of some controversy, but a typical set of values is%
\ref\takeuchi{I thank T.~Takeuchi for providing these numbers from a
recent compilation by P.~Langacker.}
\eqn\stuvalues{\eqalign{
&S= -0.8 \pm 0.5 \ts,\cr
&T= -0.2 \pm 0.4 \ts,\cr
&U= +0.2 \pm 0.9 \ts.\cr}}

The contributions to $S$ that arise in various models of technicolor have
been estimated in the papers of Ref.~\pettests. In calculating these
contributions, it has been assumed that the strong technicolor interaction
is QCD--like. That is, it has been assumed that (1)~techni--isospin (measured
by $M_{T_U} - M_{T_D}$) is a good approximate symmetry; (2)~the chiral
perturbation expansion is accurate for technipions; (3)~the spectrum of
technihadrons may be scaled from QCD as we did, e.g., in Eq.~\thadrons;
(4)~asymptotic freedom sets in rapidly above $\Lambda_{TC}$, so that
Weinberg's spectral function sum rules converge rapidly;%
\ref\sfsr{S.~Weinberg, {\it Phys.~Rev.~Lett.}~{\bf 18} (1967) 507\semi
K.~G.~Wilson, {\it Phys.~Rev.}~{\bf 179} (1969) 1499\semi
C.~Bernard, A.~Duncan, J.~Lo~Secco and S.~Weinberg, {\it Phys.~Rev.}~{\bf
D12} (1975) 792.}
and (5)~vector--meson dominance of spectral functions is valid, i.e., they
are saturated by vector axial--vector meson poles, typically the lowest
lying.

If techni--isospin is a good symmetry, then $S$ may be written as the
following spectral integral:
\eqn\spectral{S = -4 \pi \left[ \Pi_{VV}^{'}(0) - \Pi_{AA}^{'}(0) \right]
= {1 \over {3\pi}} \int {ds \over {s}} \left [ R_V(s) - R_A(s) \right]
\ts.}
Here, $R_V$ and $R_A$ are the analogs for the weak--isospin vector and
axial--vector currents of the classic ratio of cross sections, $R(s) =
\sigma(e^+e^- \ra {\rm hadrons}) / \sigma(e^+e^- \ra \mu^+ \mu^+)$. Peskin
and Takeuchi\cite{\pettests}\ used QCD as an analog computer (vector meson
dominance applied to spectral function sum rules) to estimate this
integral. In the narrow--resonance approximation, their result is
\eqn\pesktak{
S = 4 \pi \left(1 + {M^2_{\rho_T} \over{M^2_{a_{1T}}}}\right ) {F^2_\pi \over
{M^2_{\rho_T}}} \simeq 0.25 \ts N_D \ts {N_{TC}\over{3}} \ts.}
In the second equality, the scaling formula, Eq.~\thadrons, was used.

Golden and Randall, Holdom and Terning, and others\cite{\pettests}\
estimated the leading chiral--logarithmic contribution, $S_\chi$, to $S$ of
the technipions that occur in occur in multi--doublet technicolor models.
Their result is
\eqn\goldrand{
S > S_\chi \simeq {1 \over {12 \pi}} (N_D^2 -1) \log\left({M^2_{\rho_T}
\over{M^2_{\pi_T}}} \right)\simeq 0.08 (N_D^2-1) \ts.}

These estimates agree for the popular choice of the one--family model,
$N_D=N_{TC}=4$, in which case $S \simeq 1$, approximately four standard
deviations away from the central value quoted above.%
\ref\cdg{However, see
R.~S.~Chivukula, M.~Dugan and M.~Golden, {\it Phys.~Lett.}~{\bf 292B},
(1992) 435. This paper argues convincingly that this agreement is
fortuitous.}
\nref\Oz{Coroner, Munchkin City, Land of Oz, in {\it The Wizard of Oz},
(1939).}%
Thus, to paraphrase Ref.~\Oz, ``technicolor is not only really very dead,
it's really most sincerely dead!''

\newsec{WALKING TECHNICOLOR}

The FCNC and STU difficulties of technicolor have a common cause: the
assumption that technicolor is a just a scaled--up version of QCD. This
assumption implies that asymptotic freedom sets in quickly and $\condetc
\simeq \condtc$ and, hence, the estimate $\Lambda_{ETC} \simeq 1-100\,\tev$
in Eq.~\lametc. It also means that the technihadron spectrum is just a
magnified image of the QCD--hadron spectrum, hence that $S$ is too large for
all technicolor models except, possibly, the minimal one--doublet model with
$N_{TC} = 2$~or~3. Therefore, it may be possible to find a solution to
these difficulties in a technicolor theory whose gauge dynamics are
distinctly not QCD--like. A technicolor theory in which the gauge coupling
evolves slowly---``walks''---is so far the only promising example of
this.%
\ref\wtc{B.~Holdom, {\it Phys.~Rev.}~{\bf D24} (1981) 1441;
{\it Phys.~Lett.}~{\bf 150B} (1985) 301\semi
T.~Appelquist, D.~Karabali and L.~C.~R. Wijewardhana, {\it Phys.~Rev.~Lett.}~
{\bf 57} (1986) 957\semi
T.~Appelquist and L.~C.~R.~Wijewardhana, {\it Phys.~Rev.}~{\bf D36}
(1987) 568\semi
K.~Yamawaki, M.~Bando and K.~Matumoto, {\it Phys.~Rev.~Lett.}~{\bf 56}
(1986) 1335\semi
T.~Akiba and T.~Yanagida, {\it Phys.~Lett.}~{\bf 169B} (1986) 432.}
This section presents a short introduction to walking technicolor. It is
hoped that it will whet the reader's appetite for more in--depth study.
Strong extended technicolor---invoked to achieve a large enough top--quark
mass---is discussed here in only the briefest terms.%
\nref\setca{T.~Appelquist, T.~Takeuchi, M.~B.~Einhorn, L.~C.~R.~Wijewardhana,
{\it Phys.~Lett.} {\bf 220B} (1989) 223 \semi
T.~Takeuchi, Phys. Rev. {\bf D40} (1989) 2697 \semi
V.~A.~Miransky and K.~Yamawaki, {\it Mod.~Phys.~Lett.}~{\bf A4}
(1989) 129.}%
\nref\setcb{R.~S.~Chivukula, A.~G.~Cohen and K.~Lane, {\it
Nucl.~Phys.~}~{\bf B343}, 554 ( 1990).}%
\cite{\setca,\setcb}

\medskip
\noindent{\it 3.1 FCNC and STU}
\smallskip

Thus far the condensates $\condetc$ and $\condtc$ have been approximately
equal because we have assumed that $\gamma_m(\mu) \approx 3 \ts C_2(R) \ts
\alpha_{TC}(\mu)/2 \pi \ll 1$ for scales $\mu \simge \Lambda_{TC}$. The
question we must ask now is: Can $\gmm$ be large? The answer is yes---so
long as $\atc(\mu)$ is large. In that case, the perturbative formula~\gamm\
is no longer correct. An approximate expression for a nonperturbative
$\gmm$ can be obtained by solving the Schwinger--Dyson ``gap'' equation for
the technifermion's dynamical mass function, $\Sigma(p)$ in the so--called
ladder approximation. In this approximation, the
technifermion--anti-technifermion scattering kernel is given by the
one--technigluon exchange graph. The gap equation is
\eqn\gap{\Sigma(p) = 3 C_2(R)\ts \int {d^4 k \over {(2 \pi)^4}}
\ts {\atc((k-p)^2) \over {(k-p)^2}} \ts {\Sigma(k) \over {k^2}} \ts.}
The further approximation of linearizing the integral equation has been
made in Eq.~\gap. (See Ref.~\wtc\ for details.) {\it If} the gauge
coupling $\atc$ runs {\it slowly} over the entire range of momenta for
which the mass function $\Sigma$ is appreciable, it is a good approximation
to take it outside the integral. The solution for $\Sigma$ then has the
form%
\ref\footh{Another way to obtain Eq.~\sigmap\ is to use the operator product
expansion for $\Sigma(p)$ at the large momenta where the linearized gap
equation is valid; see Ref.~\kdlhdp. The result for technifermions $T$
is $\Sigma(p) \simeq (3 C_2(R) \ts \atc(p)/\pi) \ts \langle \ol T T \rangle_p
/p^2$.}
\eqn\sigmap{\Sigma(p) \simeq \Sigma(0) \left({\Sigma(0)\over{p}}\right)^{2 -
\gmm(\mu)} \ts,}
where
\eqn\nonpert{\eqalign{
&\gmm(\mu) = 1 - \sqrt{1-\atc(\mu)/\alpha_C} \ts ;\cr
&\alpha_C  = {\pi \over{3 C_2(R)}} \ts\ts.\cr}}
This reduces to the perturbative result~\gamm\ in the small--coupling
limit.%
\ref\footi{The $O(\atc)$ ladder equation, \gap, is written in the Landau gauge.
However, the anomalous dimension $\gmm$ in Eq.~\nonpert\ is
gauge--invariant---to this order in $\atc$.}

The ladder approximation indicates that spontaneous chiral symmetry
breaking occurs if and only if $\atc$ reaches the ``critical coupling''
$\alpha_C$, given in the ladder approximation by $\pi/3 C_2(R)$.%
\nref\acrit{A.~G.~Cohen and H.~Georgi, {\it Nucl.~Phys.~}~{\bf B314}
(1989) 7\semi U.~Mahanta, {\it Phys.~Rev.~Lett.}~{\bf 62} (1989) 2349.}%
\nref\ladder{T.~Appelquist, K.~Lane and U.~Mahanta, {\it Phys.~Rev.~Lett.}~{\bf
61} (1988) 1553.}%
\cite{\wtc,\acrit,\ladder} No one has yet provided a rigorous
proof of this statement, but it has become part of the accepted lore
of dynamical symmetry breaking. Thus, the chiral--symmetry breaking scale
$\Lambda_{TC}$ is {\it defined} by the condition
\eqn\chisb{\atc(\Lambda_{TC}) = \alpha_C \Longleftrightarrow
\gamma_m(\Lambda_{TC}) = 1 \ts.}
In other words, $\gamma_m(\Lambda_{TC}) = 1$ is the signal for chiral
symmetry breakdown. Cohen and Georgi, and Mahanta, Ref.~\acrit, have argued
that this is the correct {\it nonperturbative} signal.

In QCD, the lore then goes, the anomalous dimension $\gmm$ of the quark
bilinear, $\ol q q$, starts out at unity at the quark's chiral--symmetry
breaking scale, but quickly falls to its perturbative value
$2\ts \alpha_{QCD}(\mu)/\pi$ above $\mu \simeq 1\,\gev$. This is precocious
asymptotic freedom. To keep $\gmm$ large in technicolor, we must require
that the $G_{TC}$ $\beta$--function is small above $\Lambda_{TC}$:
%and remains small for a {\it large} range of momenta above this scale:
%
\eqn\betatc{
\beta(\atc(\mu)) = \mu \ts {d \atc(\mu) \over {d \mu}}
\simeq 0 \ts\ts {\rm for} \ts\ts \Lambda_{TC} < \mu < \Lambda \ts,}
where $\Lambda \gg \Lambda_{TC}$. Thus, the technicolor coupling walks
rather than runs. We do not know whether it is possible to construct a
walking gauge theory. The question is essentially nonperturbative. If the
$\beta$--function is made small through a given low order of perturbation,
one can never be sure that its higher order terms are negligible.

Finally, if $\beta(\atc) \simeq 0$ all the way to the ETC scale, so that
$\gmm(\mu) \cong 1$ for $\Lambda_{TC} < \mu < \Metc$, then hard fermion
masses are given by%
\ref\footj{The beta function cannot be zero and $\gmm = 1$ over the wide
range $\Lambda_{TC}$ to $\Metc$. This contradicts the notion of a
well--defined chiral symmetry breaking scale. Thus, Eqs.~\hard\ and
\bigpimass\ are upper limits to the mass enhancements in walking
technicolor.}
\eqn\hard{m_{q,\ell,T}(M_{ETC}) \simeq {g^2_{ETC} \over {M^2_{ETC}}} \times
\left(\condetc \cong \condtc \ts {M_{ETC} \over {\Lambda_{TC}}} \right)
\ts.}
To be quantitative, suppose
that $\Lambda_{ETC} = M_{ETC}/g_{ETC} \simeq 100 - 1000\,\tev$ and that
$\condtc \simeq 4 \pi F^3_\pi/N_D^{3/2}$, with $\Lambda_{TC} \simeq
1\,\tev$. Then, hard masses evaluated at the ETC scale are given roughly by
\eqn\rough{m_{q,\ell,T}(M_{ETC}) \simeq (0.2 - 2.0)/N_D^{3/2}\,\gev \ts.}
This enhancement is enough to accomodate the charmed quark. That may be all
we need. The strong constraint on $\Metc$ from $\vert \Delta S \vert = 2$
interactions affects masses and mixing angles in the first two generations.
The third generation masses, on the other hand,
may be associated with a considerably lower ETC
scale. To produce $m_t = 150\,\gev$, however, we would need $\Metc \simeq
F_{\pi_T}^2/m_t \simeq 500/N_D\,\gev$. We will have more to say about this
later.

Technipion masses are also enhanced by walking. The maximum ETC
contribution to the mass expected in walking technicolor is
\eqn\bigpimass{\eqalign{
(M_{\pi_T})_{ETC} &=  {\sqrt{m_T(M_{ETC}) \condetc} \over {F_{\pi_T}}}
\cr\cr
&= {g_{ETC} \over {\Lambda_{TC}}} \ts {4 \pi \Few \over {N_D}} \simeq
750/N_D\,\gev \ts. \cr}}
In this maximal case, the ETC breaking of technifermion chiral symmetry is
not small and the technipions are not pseudo--Goldstone bosons.

The calculations of the electroweak parameter $S$ described above are
suspect in a walking technicolor model. The main assumptions
of the Peskin-Takeuchi calculation\cite{\pettests}\ were that
(1)~techni--isospin is a good symmetry; (2)~Weinberg's spectral
function sum rules are valid; (3)~the spectral functions are
saturated by the lowest--lying vector and axial--vector resonances (i.e.,
vector--meson dominance); and (4)~the masses and couplings of these
mesons can be determined by scaling from QCD.

Leave the question of techni--isospin aside for a moment. Then, while the
Weinberg sum rules necessarily are valid for the weak currents, the
integral for the second sum rule converges much more slowly (to zero) in a
walking technicolor theory. This is because of the slower fall--off of the
mass function $\Sigma(p)$ compared to its QCD behavior of $1/p^2$. Thus,
the spectral functions $R_V$ and $R_A$ cannot be saturated by only the
lowest resonances. The spectral weight of $R_V(s) - R_A(s)$ is shifted to
higher $s$--values. Unfortunately, it is not possible to conclude from this
that $S$ is decreased. No one knows what the spectrum of a walking
technicolor theory looks like. (Could it be a tower of many resonances, all
contributing substantially to the spectral functions?) Thus, scaling meson
masses and parameters---even for the lowest states---is questionable.
Indeed, Ref.~\cdg\ shows that, with QCD--like dynamics, the naive
large--$N_{TC}$ scaling of the technihadron masses is invalid when there are
several technidoublets.

Return now to the issue of techni--isospin. Its breaking in ETC theories
must be fairly large to account for the $m_t - m_b$ splitting. It is known
that isospin breaking among ordinary fermions can lead to a negative $S$
and several authors have argued that this happens in walking technicolor
theories.%
\ref\isobreak{M.~Luty and R.~Sundrum, unpublished\semi
T.~Appelquist and J.~Terning, {\it Phys.~Lett.}~{\bf B315} (1993) 139\semi
B.~Balaji and K.~Lane, unpublished.}
However, one must then worry about the value of $T$. Appelquist and Terning
have argued that $S$ can be negative while $T$ remains small in a theory in
which electroweak symmetry is broken in a techni--isospin conserving way at
a high scale (this breaking accounts for most of $M_W$ and $M_Z$ and,
therefore, $\rho = 1 + \alpha T$), while the
techni--isospin is relatively badly broken at a much lower scale (where the
dominant contribution to the integral for $S$, Eq.~\spectral, comes from).

Finally, the chiral Lagrangian estimate of a lower bound for $S$ from
technipions is also likely to be unreliable. As we have just seen, chiral
perturbation theory may break down in walking technicolor theories so that
the technipions may not be properly regarded as pseudo--Goldstone bosons.

\medskip
\noindent{\it The Top--Quark Mass and Strong Extended Technicolor}
\smallskip

We noted above that an ETC scale of 1~TeV or less is required to produce a
mass of 150~GeV for the top quark, even if the maximal enhancement of
walking technicolor is invoked. This is too close to the technicolor scale,
$\Lambda_{TC}$, for the effective four--fermion interaction
$(g^2_{ETC}/\Metc^2) \ts \ol t \ts \Gamma'_\mu \ts T \ts\ts \ol T \ts
\Gamma^\mu \ts t$ to make sense. To overcome this obstacle, we need, in
effect, to enhance the technifermion condensate $\condt$ with an anomalous
dimension greater than one.\cite{\setca} This is not possible;
$\gmm = 1$ is its maximum value. A greater value of $\gmm$ is nonsensical
because one then has a dynamical mass $m(p)$ that falls off slower than
$1/p$. This corresponds to an explicit, hard mass in the Lagrangian. This
is discussed by Cohen and Georgi (Ref.~\acrit) and in Ref.~\setcb.

To generate a large $m_t$, the authors of Ref.~\setca\ proposed that ETC
interactions are strong enough to participate in driving the breakdown of
electroweak symmetry. That is, the four--technifermion interactions
generated by ETC--boson exchange at a scale $\Lambda_{ETC} \gg \Lambda_{TC}$
are not negligible compared to technicolor interactions {\it at the lower
scale}.%
\ref\footk{Here we are using $\Lambda_{ETC}$ to mean the scale at which ETC
interactions become strong and this is not necessarily the same as
$M_{ETC}/g_{ETC}$. For this strong--ETC scenario to make sense, we must
have $\Lambda_{ETC} \simge M_{ETC}$. Otherwise, the ETC gauge bosons
decouple from $\beta(g_{ETC})$, the ETC coupling stops
growing at $M_{ETC}$, and it never becomes strong.}
In order that there be a substantial separation between these two scales,
it is necessary that the chiral symmetry breaking phase transition induced
by the ETC interactions be at least second order.\cite{\setcb} Then, the
ETC--induced symmetry breaking scale is tunable; it will be $O(\Lambda_{TC})
\ll O(\Lambda_{ETC})$ if $g^2_{ETC}$ is within
$(\Lambda_{TC}/\Lambda_{ETC})^2$ of the critical ETC coupling to trigger
spontaneous symmetry breaking.%
\ref\footl{In the Nambu--Jona-Lasinio four--fermion model (see
Ref.~\nambu.), the scale $\Lambda_\chi$ of spontaneous symmetry breaking is
of order the model's cutoff, $\Lambda$---the analog of $\Metc$---unless the
interaction's coupling strength $g^2_{NJL}/4 \pi = 1 +
O(\Lambda^2_\chi/\Lambda^2)$; the critical coupling is one.}

If the ETC--generated phase transition is second order, and the ETC coupling
is tuned to be just above its critical value, there appears in the physical
spectrum a composite scalar, $\phi$.\cite{\setcb} It is formed from
technifermions bound by the ETC interactions. But, its mass $M_\phi$ is
only of $O(\Lambda_{TC})$ This scalar couples to technifermions with Yukawa
couplings that are unsuppressed by $\Lambda_{ETC}$. Hence, it develops a
vacuum expectation value of order $\condtc/M^2_\phi \simeq \Lambda_{TC}$
and this, in turn, imparts a ``hard'' mass of $O(\Lambda_{TC})$ to the
technifermions. This has the effect of $\gmm \simeq 2$, but the anomalous
dimension really is not that large. If the scalar also has unsuppressed
couplings to the top quark, then $m_t$ will also be enhanced to
$O(100\,\gev)$. This scalar's phenomenology has been shown to be consistent
with experimental limits on both FCNC%
\ref\lizfcnc{E.~H.~Simmons, {\it Nucl.~Phys.}~{\bf B312} (1989) 253.}
and oblique radiative corrections.%
\ref\lizstu{C.~D.~Carone and E.~H.~Simmons, {\it Nucl.~Phys.}~{\bf
B397} (1993) 591.}

It is very much an open question whether strong extended technicolor can
generate a top--bottom splitting of order 150~GeV.
\nref\tasetc{T.~Appelquist, T.~Takeuchi, M.~B.~Einhorn,
L.~C.~R.~Wijewardhana, {\it Phys.~Lett.} {\bf 232B} (1989) 211\semi
T.~Appelquista and O.~Shapira, {\it Phys.~Lett.}~{\bf 249B} (1990) 83\semi
T.~Appelquist, J.~Terning and L.~C.~R.~Wijewardhana, {\it
Phys.~Rev.}~{\bf D43} (1991) 646.}
Some attempts in this direction are described Ref.~\tasetc. If strong ETC
can generate such large weak--isospin breaking, then there is the concern
that it may induce too large a value for the $T$--parameter.%
\ref\rsciso{R.~S.~Chivukula, {\it Phys.~Rev.~Lett.}~{\bf 61} (1988) 2657.}
It's the same old story: No one knows how to calculate reliably in a
strongly--coupled theory for which there is no direct experimental
guidance.

\newsec{CONCLUSIONS AND ACKNOWLEDGEMENTS}

Technicolor and extended technicolor represent the most ambitious, and so
far, the most compelling attempt to understand the physics of flavor. It
has several attractive features, but it is also widely regarded as being
``ugly'' and/or too complicated. To make real progress, we will need some
very bright ideas. However, I cannot escape the feeling that what we really
need most is data. Nothing focuses the mind like interesting experimental
results. Let us hope that we get some soon. In any case, we are bound to
need the high energy and luminosity that the LHC can provide. Let us hope,
also, that that is enough.

\bigskip\bigskip

I would like to thank the organizers of TASI, especially Stuart Raby, Terry
Walker and K.~T.~Mahathappa, for inviting me to lecture and for providing
such a nice atmosphere for the school. My thanks also go to the students.
Their enthusiasm in the face of so much bad news was heartening. I am
grateful to Sekhar Chivukula and Elizabeth Simmons for reading this
manuscript. And I am, as always, thankful to my long--term collaborators,
Tom Appelquist and Estia Eichten. They never give up. This work was
supported in part by the DOE under Grant. No.~DE-FG02-91ER40676 and by the
Texas National Research Laboratory Commission under Grant No.~RGFY93-278.

\vfil\eject
\footatend\vfill\supereject\immediate\closeout\rfile\writestoppt
\baselineskip=14pt\centerline{{\bf References}}\bigskip{\frenchspacing%
\parindent=20pt\escapechar=` \input refs.tmp\vfill\eject}\nonfrenchspacing
\enddoublecolumns
\bye